%% file: paper.tex
\input harvmac.tex

\chardef\tempcat=\the\catcode`\@
\catcode`\@=11

\def\cydot{{\mathsurround=0pt$\cdot$}}

\def\ubar#1{\oalign{#1\crcr\hidewidth
    \vbox to.2ex{\hbox{\char22}\vss}\hidewidth}}

\def\cprime{\/{\mathsurround=0pt$'$}}
\def\Cprime{{\mathsurround=0pt$'$}}
\def\cdprime{\/{\mathsurround=0pt$''$}}
\def\Cdprime{{\mathsurround=0pt$\ubar{\hbox{$''$}}$}}

\def\@gobble#1{}
\def\@testgrave{\`}
\def\@stressit{\futurelet\chartest\@stresschar }

\def\@stresschar#1{%
  \ifx #1y\def\result{\futurelet\chartest\@yligature}%
  \else \ifx #1Y\def\result{\futurelet\chartest\@Yligature}%
  \else \ifx\chartest\@testgrave \def\result{\accent"26 }%
  \else \def\result{\accent"26 #1}%
  \fi \fi \fi
  \result }

\def\@yligature{%
  \ifx a\chartest \def\result{\accent"26 \char"1F \@gobble}%
  \else \ifx u\chartest \def\result{\accent"26 \char"18 \@gobble}%
  \else \def\result{\accent"26 y}%
  \fi \fi
  \result }

\def\@Yligature{%
  \ifx a\chartest \def\result{\accent"26 \char"17 \@gobble}%
  \else \ifx A\chartest \def\result{\accent"26 \char"17 \@gobble}%
  \else \ifx u\chartest \def\result{\accent"26 \char"10 \@gobble}%
  \else \ifx U\chartest \def\result{\accent"26 \char"10 \@gobble}%
  \else \def\result{\accent"26 Y}%
  \fi \fi \fi \fi
  \result }

\def\!{\ifmmode \mskip-\thinmuskip \fi}
\def\cyracc{%
  \def\cydot{{\kern0pt}}%
  \def\cprime{\char"7E }\def\Cprime{\char"5E }%
  \def\cdprime{\char"7F }\def\Cdprime{\char"5F }%
  \def\dbar{dj}\def\Dbar{Dj}%
  \def\dz{\char"1E }\def\Dz{\char"16 }%
  \def\dzh{\char"0A }\def\Dzh{\char"02 }%
  \def\'##1{\if c##1\char"0F %
    \else \if C##1\char"07 %
    \else \accent"26 ##1\fi \fi }%
  \def\`##1{\if e##1\char"0B %
    \else \if E##1\char"03 %
    \else \errmessage{accent \string\` not defined in cyrillic}%
        ##1\fi \fi }%
  \def\=##1{\if e##1\char"0D %
    \else \if E##1\char"05 %
    \else \if \i##1\char"0C %
    \else \if I##1\char"04 %
    \else \errmessage{accent \string\= not defined in cyrillic}%
        ##1\fi \fi \fi \fi }%
  \def\u##1{\if \i##1\accent"24 i%
    \else \accent"24 ##1\fi }%
  \def\"##1{\if \i##1\accent"20 \char"3D %
    \else \if I##1\accent"20 \char"04 %
    \else \accent"20 ##1\fi \fi }%
  \def\!{\ifmmode \def\result{\mskip-\thinmuskip}%
    \else \def\result{\@stressit}\fi \result}}

\def\keep@cyracc{\let\cyr=\relax \let\i=\relax
        \let\ubar=\relax \let\cydot=\relax
        \let\cprime=\relax \let\Cprime=\relax
        \let\cdprime=\relax \let\Cdprime=\relax
        \let\dbar=\relax \let\Dbar=\relax
        \let\dz=\relax \let\Dz=\relax
        \let\dzh=\relax \let\Dzh=\relax
        \let\'=\relax \let\`=\relax \let\==\relax
        \let\u=\relax \let\"=\relax \let\!=\relax }

\catcode`\@=\tempcat
% \endinput
 \newfam\cyrfam
 \font\tencyr=wncyr10
\def\cyr{\fam\cyrfam\tencyr\cyracc}

\Title{\vbox{\baselineskip12pt
\hbox{hep-th/9906235}
\hbox{ITEP-TH-36/96}
\hbox{HUTP-97/A099}}}
{\vbox{
\centerline{Duality in Integrable Systems and Gauge Theories}
}}
\vskip 0.2cm
\centerline{V.~ Fock$^{1}$, A.~ Gorsky$^{1}$, N.~ Nekrasov$^{1,3}$, V.~
Rubtsov$^{1,2}$}
\centerline{\cyr $^{1,2,3}$ Institut Teoretiqesko\u\i\quad i
E1ksperimentalp1no\u\i\quad Fiziki, 117259, Moskva, Rossiya} \centerline{\rm
$^{3}$ D\'epartement de Math\'ematiques, Universit\'e d'Angers, 49045,
Angers, France} \centerline{$^{3}$ Lyman Laboratory of Physics, Harvard
University, Cambridge MA 02138, USA} \vskip 1cm \centerline{\tt fock,
gorsky@vitep5.itep.ru, nikita@bohr.harvard.edu, volodya@orgon.univ-angers.fr}
\vskip 0.2cm
We discuss various dualities, relating integrable systems and
show that these dualities are explained in the framework
of Hamiltonian and Poisson reductions. The dualities we study
shed some light on the known integrable systems as well as allow
to construct new ones, double elliptic among them. We also discuss
applications to the (supersymmetric) gauge theories in various dimensions.
\vskip 0.1cm \Date{06/99}
%%%%%%%%%%%%%%%%%%%%

\def\sssec#1{$\underline{\rm #1}$}
\def\unlockat{\catcode`\@=11}
\def\lockat{\catcode`\@=12}
\unlockat
\def\newsec#1{\global\advance\secno by1\message{(\the\secno. #1)}
\global\subsecno=0\global\subsubsecno=0\eqnres@t\noindent
{\bf\the\secno. #1}
\writetoca{{\secsym} {#1}}\par\nobreak\medskip\nobreak}
\global\newcount\subsecno \global\subsecno=0
\def\subsec#1{\global\advance\subsecno by1\message{(\secsym\the\subsecno. #1)}
\ifnum\lastpenalty>9000\else\bigbreak\fi\global\subsubsecno=0
\noindent{\it\secsym\the\subsecno. #1}
\writetoca{\string\quad {\secsym\the\subsecno.} {#1}}
\par\nobreak\medskip\nobreak}
\global\newcount\subsubsecno \global\subsubsecno=0
\def\subsubsec#1{\global\advance\subsubsecno by1
\message{(\secsym\the\subsecno.\the\subsubsecno. #1)}
\ifnum\lastpenalty>9000\else\bigbreak\fi
\noindent\quad{\secsym\the\subsecno.\the\subsubsecno.}{#1}
\writetoca{\string\qquad{\secsym\the\subsecno.\the\subsubsecno.}{#1}}
\par\nobreak\medskip\nobreak}

\def\subsubseclab#1{\DefWarn#1\xdef #1{\noexpand\hyperref{}{subsubsection}%
{\secsym\the\subsecno.\the\subsubsecno}%
{\secsym\the\subsecno.\the\subsubsecno}}%
\writedef{#1\leftbracket#1}\wrlabeL{#1=#1}}% Macros for boxes
\lockat

%%%%%%%%%%%%%%%%%%%%%  Rublenye bukvy   %%%%%%%%%%%%%%%%%%%%%%%%
\def\IB{\relax\hbox{$\inbar\kern-.3em{\rm B}$}}
\def\IC{\relax\hbox{$\inbar\kern-.3em{\rm C}$}}
\def\ID{\relax\hbox{$\inbar\kern-.3em{\rm D}$}}
\def\IE{\relax\hbox{$\inbar\kern-.3em{\rm E}$}}
\def\IF{\relax\hbox{$\inbar\kern-.3em{\rm F}$}}
\def\IG{\relax\hbox{$\inbar\kern-.3em{\rm G}$}}
\def\IGa{\relax\hbox{${\rm I}\kern-.18em\Gamma$}}
\def\IH{\relax{\rm I\kern-.18em H}}
\def\IK{\relax{\rm I\kern-.18em K}}
\def\IL{\relax{\rm I\kern-.18em L}}
\def\IP{\relax{\rm I\kern-.18em P}}
\def\IR{\relax{\rm I\kern-.18em R}}
\def\IZ{\relax\ifmmode\mathchoice
{\hbox{\cmss Z\kern-.4em Z}}{\hbox{\cmss Z\kern-.4em Z}}
{\lower.9pt\hbox{\cmsss Z\kern-.4em Z}}
{\lower1.2pt\hbox{\cmsss Z\kern-.4em Z}}\else{\cmss Z\kern-.4em Z}\fi}
\def\ocom{\relax{\raise1.2pt\hbox{$\otimes$}\kern-.5em\lower.1pt\hbox{,}}\,}
%%%%%%%%%%%%%%%%%%%% Calligraphic letters  %%%%%%%%%%%%%%%%%%%%%%%
\def\CA {{\cal A}}

\def\CF {{\cal F}}
\def\CG {{\cal G}}
\def\CH {{\cal H}}

\def\CK {{\cal K}}
\def\CL {{\cal L}}
\def\CM {{\cal M}}
\def\CN {{\cal N}}
\def\CO {{\cal O}}

\def\CS {{\cal S}}

\def\CW {{\cal W}}
\def\CX {{\cal X}}

%%%%%%%%%%%%%%%%%%%%%%%%%% Derivatives  %%%%%%%%%%%%%%%%%%%%%%%%

\def\wwp{{\wp}_{\tau}(q)}

\def\p{\partial}

%%%%%%%%%%%%%%%%%%%% letters with bar %%%%%%%%%%%%%%%%%%%%%%%%%%

%%%%%%%%%%%%%%%%%%%%%%%%%%% Math symbols %%%%%%%%%%%%%%%%%%%%%%%

\def\Tr{\rm Tr}
\def\Id{{\rm Id}}

%%%%%%%%%%%%%%%%%%%%%%%%%%%%%%%%%%%%%%%%%%%%%%%%%%%%%%%%%%%%%%%%%
\font\manual=manfnt \def\dbend{\lower3.5pt\hbox{\manual\char127}}

\def\c{\cdot}
\def\half {{1\over 2}}
\def\ch{{\rm ch}}

\def\inbar{\,\vrule height1.5ex width.4pt depth0pt}
%%%%%%%%%%%%%%%%% Lie algebras %%%%%%%%%%%%%%%%%%%%%%

\def\lieg{{\underline{\bf g}}}

%%%%%%%%%%%%%%%%%%%%%%%%%%%%%%%%%%%%%%%%%%%%%%%%%%%%%%%%%%%%%%%%%

% Macros for boxes

\font\cmss=cmss10 \font\cmsss=cmss10 at 7pt

\def\boxit#1{\vbox{\hrule\hbox{\vrule\kern8pt
\vbox{\hbox{\kern8pt}\hbox{\vbox{#1}}\hbox{\kern8pt}}
\kern8pt\vrule}\hrule}}
\def\mathboxit#1{\vbox{\hrule\hbox{\vrule\kern8pt\vbox{\kern8pt
\hbox{$\displaystyle #1$}\kern8pt}\kern8pt\vrule}\hrule}}

%% ANOTHER SET OF MACROS

\def\inbar{\,\vrule height1.5ex width.4pt depth0pt}

\font\cmss=cmss10 \font\cmsss=cmss10 at 7pt

%REFERENCES
%
\lref\lizzidadda{
A.~D'Adda, F~.Lizzi, ``Space dimensions from supersymmetry for the
${\CN}=2$ spinning string: a four dimensional model'', Phys.~Lett.{\bf B191} (1987) 85}
\lref\matone{G.~Bonelli, M.~Matone, ``Nonperturbative Relations in N=2 SUSY Yang-Mills and WDVV
equation'', Phys.~Rev.~Lett.{\bf 77} (1996) 4712, hep-th/9605090}

\lref\landau{L.~ Landau, E.~ Lifshitz, ``Quantum Mechanics (Non-relativistic theory)'', Pergamon Press,  1977}
%%%%%%%%%%%%%%%%%%%%%%%%%%%%% Alg Geom %%%%%%%%%%%%%%%%%%%%%%%%%%%%%
%%%%%%%%%%%%%% Dubrovin %%%%%%%%%%%%%
\lref\dubrovin{B.~ Dubrovin, ``Geometry of $2d$ Topological Theories'', hep-th/9407018, SISSA-89/94-FM}

\lref\separ{A.~Gorsky, N.~Nekrasov, V.~Rubtsov, ``Separation of variables,
Hilbert Schemes and D-Branes'', hep-th/9901089}

%%%%%%%%%%%%%%%%%% Vafa %%%%%%%%%%%%%%%%%%%%%%%%%%%%%%%
\lref\vafarev{C.~ Vafa, ``Lectures on strings and dualities'', hep-th/9702201}
\lref\ogvf{H. Ooguri and C. Vafa, ``Self-Duality
and ${\CN} = 2$ String Magic,'' Mod.Phys.Lett. {\bf A5} (1990) 1389-1398\semi
``Geometry
of ${\CN} = 2 $ Strings,'' Nucl.Phys. {\bf B361}  (1991) 469-518.}
%%%%%%%%%%%%%%%%%% Witten %%%%%%%%%%%%%%%%%%%%%%%%%
\lref\WitDonagi{R.~ Donagi, E.~ Witten,
``Supersymmetric Yang-Mills Theory and
Integrable Systems'', hep-th/9510101, Nucl.Phys.{\bf B}460 (1996) 299-334}

\lref\Witdgt{ E.~ Witten, ``On Quantum gauge theories in two
dimensions,'' CMP {\bf  141}  (1991) 153\semi
 ``Two dimensional gauge
theories revisited'', J. Geom. Phys. {\bf 9} (1992) 303-368}

\lref\witseiii{N.~Seiberg, E.~Witten,
``Gauge Dynamics and Compactification to Three Dimensions'', hep-th/9607163}

\lref\witdynii{E.~Witten, ``Some Comments on String Dynamics'',
hep-th/9507121}

\lref\bpz{A.A. Belavin, A.M. Polyakov, A.B. Zamolodchikov,
``Infinite conformal symmetry in two-dimensional quantum
field theory,'' Nucl.Phys.B241:333,1984}

\lref\hi{N.~ Hitchin, ``Stable Bundles And Integrable Systems'', Duke Math. Journal, Vol. 54, No. 1 (1987)}
\lref\hid{N.~ Hitchin, ``The Self-Duality Equations On A Riemann Surface'', Proc. London Math. Soc. (3) {\bf 55}(1987) 59-126}

\lref\emilnickone{E. Martinec and N. Warner, ``Integrable systems and
supersymmetric gauge theory'', hep-th/9509161, Nucl. Phys. {\bf B459}, 97}

%%%%%%%%%%%%%%% Faddeev, Novikov %%%%%%%%%%%%%%%%%%%%%%%%%%%%%%
\lref\fs{L.~ Faddeev and S.~ Shatashvili, Theor. Math. Fiz., 60 (1984)
206}
\lref\faddeevlmp{L.D.~ Faddeev, ``Some Comments on Many Dimensional
Solitons'',
Lett. Math. Phys., 1 (1976) 289-293.}
\lref\nov{ S. Novikov,"The Hamiltonian
formalism and many-valued analogue of  Morse theory",
Russian Math.Surveys 37:5(1982),1-56}
\lref\fsi{L.~ Faddeev, Phys. Lett. B145 (1984) 81.}
\lref\fadba{L.~ Faddeev, ``How Bethe Ansatz Works for Integrable Model'', 1995 Les-Houches lecture notes, hep-th/9605187}
\lref\novves{S.~Novikov, A.~Veselov, Proc. Steklov Math. Institute, {\bf 3}
(1985) 53-65 }
%%%%%%%% I. Frenkel and E. Frenkel %%%%%%%%%%%
\lref\efren{P.~Etingof, I.~Frenkel, ``Central Extensions of Current Groups
in Two Dimensions'', CMP {\bf 165} (1994) 429-444}
\lref\efk{P.~ Etingof, I.~Frenkel, A.~Kirillov, Jr.,
``Spherical functions on affine Lie groups'', hep-th/9403168}
\lref\EdFr{E.~ Frenkel, ``Affine algebras,
Langlands Duality and Bethe Ansatz'', in Proceeding of {\bf XI} ICMP
(Paris, 1994), International Press, Cambridge MA 1995; q-alg/9506003}
\lref\afc{G.~Arutyunov , S.~Frolov and P.~Medvedev
``Elliptic Ruijsenaars-Schneider model from the cotangent bundle
over the two-dimensional current group'', hep-th/9608013 , J.~Math.~Phys.
38 (1997) 5682 \semi
``Elliptic Ruijsenaars-Schneider model via the Poisson reduction
of the affine Heisenberg double'', hep-th/9607213 , J.~Phys.~{\bf A} (1997)
505}

%%%%%%%%%%%%%%%%%%%% Shifman %%%%%%%%%%%%%%%%%%%%%%%%%%%
\lref\ShiBeta{V.~ Novikov, M.A.~Shifman, A.I.~Vainshtein, V.I.~Zakharov,
``Exact  Gell-Mann-Low Function of Supersymmetric Yang-Mills
Theories From Instanton
Calculus'', Nucl.Phys.{\bf B} 229 (1983) 381\semi
``Beta Function in Supersymmetric Gauge Theories:
Instantons Versus Traditional Approach'', Phys.Lett.{\bf B} 166
(1986) 329}
\lref\fr{V.~ Fock, A.~ Rosly, ``Flat connections and
polyubles'', Teor.  Math. Fiz. {\bf 95} (1993) 2, 228-238.}
\lref\SeObserv{N.~Seiberg, Nucl. Phys. {\bf B303} (1988) 286.}
\lref\SeBeta{N.~Seiberg, ``Supersymmetry and
Non-perturbative Beta-Functions'', Phys.~Lett.{\bf B}206 (1988) 75}
\lref\SeWi{N.~Seiberg, E.~Witten, ``Electric-Magnetic Duality,
Monopole Condensation, And Confinement in ${\CN}=2$ Supersymmetric
Yang-Mills Theory ''
Nucl. Phys. {\bf B}426 (1994) 19-52 (and erratum - ibid. {\bf B}430 (1994)
485-486 )\semi ``Monopoles, Duality and Chiral Symmetry Breaking in ${\CN}=2$
Supersymmetric QCD'', hep-th/9408099, Nucl. Phys. {\bf B}431 (1994) 484-550.}
\lref\seiberg{
K.~ Intriligator and N.~ Seiberg,
``Phases of ${\CN}=1$ supersymmetric gauge theories in four dimensions'',
hep-th/9408155\semi
K.~ Intriligator, N.~ Seiberg, and S.~ H.~ Shenker,
``Proposal for a Simple Model of Dynamical SUSY Breaking'',
hep-ph/9410203\semi
 N.~Seiberg, ``Electric-Magnetic Duality in Supersymmetric Non-Abelian Gauge
Theories'', hep-th/9411149\semi
K.~Intriligator and N.~Seiberg
``Duality, Monopoles, Dyons, Confinement
and Oblique
Confinement in Supersymmetric $SO(N_c)$ Gauge Theories'',
 hep-th/9503179}
\lref\GaMorSe{O.~Ganor, D.~Morrison, N.~Seiberg, ``Branes, Calabi-Yau Spaces and Compactification of
Six Dimensional $E_{8}$ Theory'', hep-th/9610251}
\lref\Seiii{K.~Intrilligator, N.~Seiberg, ``Mirror
symmetry in three dimensional gauge theories'', hep-th/9607207,
Phys. Lett. {\bf B}387 (1996) 513}
\lref\Kapiii{A.~Kapustin, ``Solution of ${\CN}=2$ gauge theories via
compactification to three dimensions'', hep-th/9804069,
Nucl. Phys. {\bf B}534 (1998) 531}
\lref\DouProbe{M.~ Douglas, ``Gauge Fields and $D$-branes'', hept-h/9604198}
\lref\doureview{M.~ Douglas, ``Superstring Dualities, Dirichlet Branes and the Small Scale
Structure of Space-Time'', hep-th/9610041}
\lref\argfar{P.C.~ Argyres, A.E.~ Farragi,
    ``The Vacuum Structure and Spectrum
        of ${\CN}=2$
        Supersymmetric $SU(N)$ Gauge Theory'',
hep-th/9411057\semi A.~ Klemm, W.~ Lerche, S.~ Theisen and S.~ Yankielowicz,
        ``Simple Singularities and
        ${\CN}=2$ Supersymmetric Yang-Mills Theory'', hep-th/9411048}
\lref\jpower{J.~Schwarz, ``The Power of $M$-Theory'', Phys. Lett. {\bf B377} (1996) hep-th/9508143}
\lref\gncal{A.~ Gorsky, N.~ Nekrasov, ``Hamiltonian
systems of Calogero type and Two Dimensional Yang-Mills
Theory'', Nucl. Phys. {\bf B} 414 (1994), 213-238, hep-th/9304047}
\lref\gnell{A.~ Gorsky, N.~ Nekrasov, ``Elliptic Calogero-Moser System from
Two Dimensional Current Algebra'', hep-th/9401021}
\lref\gnru{A.~ Gorsky, N.~ Nekrasov, ``Relativistic Calogero-Moser model as
gauged WZW theory'', hep-th/9401017, Nucl.Phys. {\bf B} 436 (1995) 582-608}
\lref\nhol{N.~ Nekrasov,    ``Holomorphic bundles and many-body
systems'', hep-th/9503157, CMP{\bf 180} (1996),  587-603}
\lref\fivedim{N.~ Nekrasov, ``Five dimensional gauge theories and relativistic
integrable systems'', hep-th/9609219, Nucl. Phys. {\bf B} 531 (1998) 323-344}
\lref\gmmra{A.~ Gorsky,~ A.~ Marshakov,~A.~ Mironov, A.~ Morozov,
``${\CN}=2$ Supersymmetric QCD and Integrable Spin chains:
Rational Case $N_{f} < 2N_{c}$ '', hep-th/9603140, Phys.Lett. {\bf B} 380
(1996) 75}
\lref\ggmone{A. Gorsky, S.Gukov and A.Mironov,
``Multiscale ${\CN}=2$ susy field
theories, integrable systems and their stringy/brane origin'',
hep-th/9707120, Nucl. Phys. {\bf B} 517 (1998) 409 \semi
``Integrable spin chains and supersymmetric Yang-Mills theories in higher dimensions'',
hep-th/9710239, Nucl.Phys. {\bf B} 518 (1998) 683}
\lref\gmmm{A.~ Gorsky,~ I.~ Krichever,~ A.~ Marshakov,~
A.~Morozov,~ A.~ Mironov,~ ``Integrablity and Seiberg-Witten Exact Solution'',
hep-th/9505035,  Phys. Lett. {\bf B}355 (1995) 466-474}
\lref\bbkt{O.~ Babelon, E.~ Billey, I.~ Krichever, M.~ Talon,
``Spin generalization of
the Calogero-Moser system and the Matrix KP equation'',
hep-th/9411160}
\lref\kr{I.~ Krichever, Funk. Anal. and Appl., {\bf 12}
(1978),  1, 76-78; {\bf 14} (1980), 282-290}
\lref\krz{I.~ Krichever, A.~ Zabrodin, ``Spin Generalization of
Ruijsenaars-Schneider model, Non-abelian Toda Chain and Representations
of Sklyanin Algebra'', hep-th/9505039}
\lref\witvar{E. Witten, ``String theory dynamics in various dimensions'',
hep-th/9503124, Nucl.Phys. {\bf B} 443 (1995), 85}
\lref\joed{J.~ Polchinski, ``Dirichlet-branes and Ramond-Ramond charges'',
hep-th/9510017, Phys.Rev.Lett. {\bf 75}, 4724}
\lref\jorev{J.~  Polchinski, ``TASI lectures on $D$-branes'', hep-th/9611050}
\lref\beil{A.~Beilinson, V.~Drinfeld,
``Quantization of Hitchin's fibration and Langlands' program'', in
  `Algebraic and geometric methods in mathematical physics' (Kaciveli, 1993), 3--7,
  Math. Phys. Stud., 19, Kluwer Acad. Publ., Dordrecht, 1996.}
\lref\kzb{V.~Knizhnik, A.~Zamolodchikov, Nucl. Phys. {\bf B} 247 (1984)
83-103\semi
D.~ Bernard, Nucl.  Phys.  {\bf B}303 (1988) 77, {\bf B}309
(1988), 14}
\lref\amm{H.~Airault, H.~McKean and Y.~Moser, Comm. on Pure and
Appl.  Math., vol. {\bf XXX}, (1977), 95-148}
\lref\w{G.~ Wilson,
``Bispectral Commutative Ordinary Differential Operators'', J. reine angev.
Math. {\bf 442} (1993) 177-204} \lref\kas{A.~Kasman, ``Bispectral KP solution
and linearization of Calogero-Moser particle systems'',  CMP {\bf 172}
(1995), 427-448} \lref\c{F.~ Calogero, J.Math.Phys. {\bf 12} (1971) 419}
\lref\ch{I.~ Cherednik, ``Difference-elliptic operators and root systems'',
hep-th/9410188} \lref\eq{P.~Etingof, ``Quantum integrable systems and
representations of Lie algebras'', hep-th/9311132}
\lref\ekq{P.~Etingof, A.~Kirillov, Jr., ``Macdonald polynomials and
representations of quantum group', hep-th/9312103\semi
``Representation-theoretic proof of the inner product and symmetry
identities for Macdonald's polynomials'', hep-th/9410169}
\lref\ek{P.~Etingof, A.~Kirillov, Jr., ``On the affine analogue of Jack's and
Macdonald's polynomials'', hep-th/9403168}
\lref\ga{R.  Garnier, ``Sur
une classe de system\`es differentiels Abeli\'ens d\'eduits de la theorie des
\'equations lin\'eares'', Rend. del Circ.  Matematice Di Palermo, {\bf 43},
vol. 4 (1919)}
\lref\gau{M. Gaudin, Jour.  Physique, {\bf 37} (1976),
1087-1098}
\lref\ivi{D.~ Ivanov, `` Knizhnik-Zamolodchikov-Bernard equations
on Riemann surfaces'', hep-th/9410091\semi
``Knizhnik-Zamolodchikov-Bernard equations as quantization of a nonstationary
Hitchin system'', hep-th/9610207}
\lref\kks{D.~ Kazhdan, B.~ Kostant and S.~
Sternberg, Comm. on Pure and Appl. Math., vol. {\bf XXXI}, (1978), 481-507}
\lref\lo{A.~ Losev, ``Coset construction and Bernard Equations'', CERN-TH.6215/91}
\lref\m{J.~Moser, Adv.Math. {\bf 16} (1975), 197-220; }
\lref\rs{S.N.M. Ruijsenaars, H. Schneider, ``A New Class of Integrable
Systems and Its Relation to Solitons'',
Ann. of Physics {\bf 170} (1986), 370-405}
\lref\r{S. ~ Ruijsenaars, ``Complete Integrability of Relativistic Calogero-Moser Systems and
Elliptic Function Identities'', CMP {\bf 110}  (1987), 191-213 }
\lref\raa{S. ~ Ruijsenaars, ``Action-Angle Maps and Scattering Theory for Some Finite-Dimensional Integrable Systems'', CMP {\bf 115}  (1988), 127-165 }
\lref\rrev{S.~ Ruijsenaars, ``Finite-Dimensional Soliton Systems'', in `Integrable
and Super-Integrable Systems', eds. B.~ Kupershmidt, World Scientific, Singapore, 1986}
\lref\ruijdu{S.~ Ruijsenaars, ``Action-Angle Maps and Scattering Theory for Some Finite-Dimensional Integrable Systems $II$, $III$'', Publ. {\bf RIMS}, Kyoto Univ {\bf 30} (1994),
865-1008 and {\bf 31} (1995), 247-353}
\lref\s{B. Sutherland, Phys. Rev. {\bf A5} (1972), 1372-1376;}
\lref\sch{L. Schlesinger, `` \"Uber eien Klasse
von Differentialsystemen
beliebiger Ordnung mit festen kritischen Punkten'',
Journal f\"ur die reine und angewandte Mathematik,
Bd. CXLI (1912), pp. 96-145}
\lref\ff{B. Feigin , E. Frenkel, N. Reshetikhin, ``Gaudin Model,
Critical Level and Bethe Ansatz'', CMP {\bf  166} (1995), 27-62}
\lref\arnold{V.~ Arnol'd, ``Mathematical methods in Classical Mechanics'', Springer-Verlag, NY 1989}
\lref\op{M.~ Olshanetsky, A.~ Perelomov, Phys. Rep. {\bf 71} (1981), 313}
\lref\opr{M.~ Olshanetsky, A.~ Perelomov, Inv. Math. {\bf 31} (1976), 93} .
\lref\perelomov{A.~ Perelomov, Comm. Math. Phys. {\bf 63} (1978) 9}
\lref\perlprep{A.~ Perelomov, Preprint ITEP {\#} 27, 1976}
\lref\perlbook{A.~ Perelomov, ``Integrable Systems of Classical Mechanics and Lie Algebras'',
Moscow,  `Nauka', 1990 [in Russian] }
\lref\donagirev{R.~ Donagi, ``Seiberg-Witten Integrable Systems'', alg-geom/9705010}
\lref\lazar{R.~Donagi, E.~Lawrence, R.~Lazarsfeld,  ``A Non-linear Deformation of the Hitchin Dynamical System'', alg-geom/9504017}
\lref\cubic{R.~Donagi, E.~Markman, ``Cubics, Integrable Systems and Calabi-Yau Threefolds'',
alg-geom/9408004}
\lref\spectral{R.~Donagi, ``Spectral Covers'', alg-geom/9505009}
\lref\donagimarkman{R.~Donagi, E.~Markman,
``Spectral Curves, Algebraicaly Completely
Integrable Hamiltonian  Systems and Moduli of Bundles'',
alg-geom/9507017}

%%%%%%%%%%%%%%%%%%%%%%%%%%%%%%%%%%%%%%%%%%%%%%%%%%%%%
\centerline{\sssec{TABLE \quad OF \quad CONTENT}}
\settabs\+3...&aaa&3.3..&aaa&3.3.3..&aaaaaaaaaaaaaaaaaaaaaaaaaaaaaaaaaaa
aaaaaaaaaaaaaaa&\cr
\bigskip\bigskip\bigskip
\bigskip
\+1.&Introduction&&&&&2\cr
\medskip
\+2. &The concepts of Duality&&&&&6\cr
\medskip
\+&&2.1& {\sl AC}: $(p,q) \to (I, \varphi)$ &&&7\cr
\smallskip
\+&&2.2& {\sl AA}: $I \to I^{D}$&&&7\cr
\smallskip
\+&&2.3& Quantum duality&&&8\cr
\medskip
\+3.&Examples of dual systems. Two particles case.&&&&&&10\cr
\medskip
\+&&3.1&Classical systems&&&10\cr
\smallskip
\+&&3.2& Quantum systems &&&16\cr
\medskip
\+4.& Duality in many-body systems&&&&&20\cr
\medskip
\+&&4.1&Examples. &&&20\cr
\smallskip
\+&&4.2& Explanations: Hamiltonian and/or Poisson reductions&&&22\cr
\medskip
\+5. &Gauge theories and duality in integrable systems&&&&&36\cr
\medskip
\+&&5.1& Old approach: many-body systems as low-dimensional &&&\cr
\+&& &gauge theories  &&&36\cr
\smallskip
\+&&5.2.& New approach: many-body systems in  supersymmetric &&&\cr
\+&& & gauge theories. &&&37\cr
\smallskip
\+&&5.3& Duality in field theories vs. duality in many-body systems &&&38\cr

\vfill\eject
\input dual1.tex

\input dual2.tex

\input dual3.tex
\input dual4.tex

\input dual5.tex

\listrefs
\bye

%% file: dual1.tex
%%%%%%%  Glava1  %%%%%%%
%%%%%%  Vvedenie  %%%%%%%%

\newsec{Introduction}

Traditionally the physical applications of
integrable systems are exhausted by  the approximations
to real dynamical systems. The configurations of the
(classical or quantum) model form the phase space which is
a manifold $M$ with the symplectic two-form $\omega$ or
Poisson bi-vector $\pi$.

The phase space might carry a natural  complex structure, such
that the symplectic form $\omega$ is a holomorphic
$(2,0)$-form, the Hamiltonians $H(p,q)$ are
the holomorphic functions and the vector fields are
holomorphic vector fields (see \hi\ for example).
The use of integrable systems as describing the evolution
in the physical models is less transparent  in this case.

The integrable systems in the holomorphic sense
entered physics approximately at the same time as the string theory
did. Particle which has a one-real-dimensional
worldline is naturally described with the help of a real phase space,
its (real) time evolution being generated by the real Hamiltonian.
The  worldsheet of a string is a complex curve, embedded into the target
space. It  senses holomorphic
geometry in many different ways. In particular, the embeddings of the
worldsheet  $\Sigma$ are governed by a two dimensional conformal  field theory
on $\Sigma$.
An example of holomorphic
integrable system relevant to the latter is
the famous Hitchin system. The phase space of this
model is the cotangent bundle to the moduli space $\CM_{G}({\Sigma})$
of  holomorphic
$G$-bundles over a Riemann surface $\Sigma$.
One can think of  Knizhnik-Zamolodchikov-Bernard
\kzb\ equations in  two dimensional WZW theories as of the non-stationary
quantum version of Hitchin system
\lo\ivi\sch\ga\gau\ff.  The r\^ole of times
is played by the complex (and hypothetically  $\CW$)-moduli
of $\Sigma$. Complex time evolution occurs
also in the models of $N=(2,2)$ strings, where
 space-time may have
$(2,2)$ signature \ogvf.

However
there exist other possibilities for
integrable
system to  encode the physical information.

In particular, the rich source of holomorphic integrable systems
is the combination of supersymmetry and duality.  It is known
for some  time  now  that the holomorphy
of certain quantities (like the  superpotential  in $\CN=1$ or
prepotential in $\CN=2$) in the supersymmetric theories
in three/four dimensions yields powerful
predictions for the behavior of the quantum theory  even
in the presence  of the non-perturbative effects
\ShiBeta, \SeBeta, \SeWi,\seiberg, \argfar, \witseiii.  In particular,
the complex structure of the moduli space of vacua in
$\CN = 4$ $3d$ gauge theories and $\CN=2$ $4d$ gauge
theories can be determined revealing
the exciting link with the special nature of the geometry of
the  phase spaces of complex integrable systems
\gmmm, \emilnickone, \WitDonagi, \witseiii.
The action variables appear as the central
charges in the BPS representations of the susy algebra \gmmm.

There exists an
approach to a class of integrable system which allows to uncover
the origin of their integrability/solvability. Namely, one realizes the system
under investigation as a projection of a simple system on a larger
phase space \opr,\op.  This idea is actually a counterpart of the
main principle behind the gauge theories - the complex dynamics of the actual
world (as far as most of the fundamental
interactions are concerned) is a projection of a somewhat simpler dynamics of
the
extended phase space.  One of the goals of the present discussion is to use
the
analogy between the two ideas and explain certain properties of integrable
systems as well as gauge theories.

We are going to study the phenomenon of {\it duality} whose
precise
definition is presented shortly. Duality is a subject of much recent
investigation in the context of
(supersymmetric) gauge theories,
in which case the duality is an involution, which maps the observables
of one theory to those of another. The duality is powerful  when the coupling
constant
in one theory
is inverse  of that in another (or more generally, when small coupling
is mapped to the strong one).
 For example,
a weakly coupled (magnetic) theory can be dual to the strongly coupled
(electric) theory thus making possible to understand the strong coupling
behavior
of the latter. In particular, it was shown by N.~Seiberg and E.~Witten \SeWi\
that using the concept of duality one can find  exact low-energy
Lagrangian of $\CN=2$,$d=4$ $SU(2)$ gauge theory.
A more fascinating recent development is that the duality connecting
weak and strong coupling regimes of one or different theories
may have a geometric origin. The most notorious example of that
is provided by
$M$-theory \witvar,\jpower.
We are going to study the dualities in integrable
systems, related to the gauge theories with the
emphasis on their geometric origin.

The study of geometry of  integrable systems also allows to understand the
origin
of certain constructions of separation of variables
\separ. The similarity of this
construction to the description of the D2-brane moduli space and its r\^ole
in the understanding the string duality makes one  hope that
both subjects - many-body integrable systems and gauge theories
(more generally, D-geometry of M.~Douglas \doureview\DouProbe)
will benefit more from  each other in the near future.

\noindent\sssec{Topics \quad left \quad beyond \quad the \quad scope \quad
of \quad the \quad paper.}
To keep the size of the paper within reasonable limits we
decided to restrict our attention with the pure many-body systems.
More or less everything we have said can be carried over to the
spin systems both of the `adjoint' \bbkt\krz\ and
`fundamental' \gmmra\ type. We don't discuss extensively the relation
of our dualities in  integrable  systems to the physics of D-branes
\joed\jorev\vafarev. Some of the results in this direction together
with the applications to the theory of separation of  variables can be
found in \separ. Also, except for the general discussion and two-body
examples we don't treat quantum case. For some results related to our
main topic see \eq\ekq\ek\beil. Realizations
of elliptic Ruijsenaars-Schneider  models via
Hamiltonian and Poisson reductions can be found in \afc.

\noindent\sssec{Organization \quad of \quad the \quad paper.} Various
concepts of
duality are discussed
 in the section $2$.
The examples of the dual systems are studied in section $3$ where mostly
two-body case is treated, both classical and quantum one. Many-body systems
are studied in the section $4$ with the explanation of the dualities
between them coming from Hamiltonian/Poisson reductions. The section $5$
is devoted to the gauge dynamics and their relation
to the integrable systems discussed so far. We discuss the geometry
of the moduli spaces of vacua of supersymmetric gauge theories
in three, four, five and six dimensions and construct a little
dictionary translating the notions of integrable systems to those
of gauge theories.

\noindent\sssec{Acknowledgements.} During the five years of the development of this project we have benefited
a lot
from the discussions with our colleagues and friends on the matters
reflected in the paper. We are grateful
to
P.~Cartier, P.~Etingof,
B.~Enriquez, B.~ Feigin, G.~Felder,
D.~Ivanov, V.~Kac, D.~Kazhdan,
B.~Kostant,
A.~Kirillov, Jr.,
A.~Losev,
A.~Rosly, S.~Ruijsenaars,
S.~Shatashvili.

We would like to express our special gratitude to J.~Harnad and
M.~Olshanetsky for their interest and scepticism which
helped us to finish this paper.

We would like to thank Institute of Theoretical Physics at Uppsala University,
Institut Mittag-Leffler
and Antti Niemi for their hospitality when this project was launched and
finished.

In addition,
A.~G.~ thanks Institute of Theoretical Physics at University
of Minnesota and  Institute of Theoretical
Physics at Bern University where the part of the work was done
for the hospitality\semi

N.~N.~ thanks CERN Theory Division and Prof.  L.~Alvarez-Gaume,
Dept. of Mathematics at MIT and Prof. V.~Guillemin, Dept. of Mathematics
Yale University and Profs. I.~Frenkel and  B.~Khesin, Dept. of Mathematics at
Brandeis University and Profs. M.~Adler, D.~Buchsbaum, P.~ van Moerbeke,
Institute of Theoretical Physics
at University of Minnesota and Profs. A.~ Vainshtein, M.~Voloshin and
M.~Shifman,
UNAM at Mexico City and Prof. A.~Turbiner,
ITP at UC Santa Barbara and Prof. D.~Gross
for their hospitality and the possibility
to present some of the results of this paper there\semi

V.~R.~ thanks CMAT Ecole Polytechniques, Universite Louis Pasteur Strasbourg I
for hospitality during preparation of the manuscript.

The preliminary results of the paper were reported at the conferences:
International Conference on
String Theory, Quantum Gravity and Integrable Systems (Alushta, 1994, 1995),
 CIRM Colloquium ``Geometrie symplectique
d'espaces de modules et systemes integrables'' (Luminy 1995),
Workshop on Symplectic Geometry (MIT, 1995),
Journee ``Systemes integrables et groupes
quantiques'' (Ecole Polytechnique Palaiseau, 1995),
33d Winter School on String Theory and Duality (Karspasz, 1997),
Workshop on Calogero-Moser-Sutherland systems (CRM, Montreal, 1997),
Journee ``Geometrie algebrique, transformationes isomonodromiques
et dualite'' (Universite d'Artois, Lens, 1998)

We thank the organizers of the conferences for giving us
the opportunity to present our
results there.

Research of A.~G.~
was supported partially by the grants INTAS-97-0103
and Schweizerisher Nationalfonds\semi
Research of N.~N.~ was supported by Princeton University through Ogden Porter Jacobus honorific fellowship, through the fellowship from
Harvard Society of Fellows, partly
by NSF under the grants PHY-92-18167, PHY-94-07194,
PHY-98-02709, partly by {\cyr RFFI}
under grant 96-02-18046\semi
N.~N. and V.~R~ are partly supported by the
grant 96-15-96455 for scientific schools\semi
A.~G~., V.~F. and V.~R.~ are partly supported by {\cyr RFFI} under grant
95-01-01110.

\vfill\eject

%% file: dual2.tex
%%%%%%% Glava 2 %%%%%%%%
%%% Dual'nost' i s chem ee edyat %%%%%
%%%%%%%%%%%%%%%%%%%%%%
%%%%%%%%%%%%%%%%%%%%%
\def\ih{{\vec{\bf h}}}
\newsec{The concepts of Duality:}

Let $(M, \omega)$ be a symplectic manifold.
There exist Darboux local coordinates (cf. \arnold ) in which
the symplectic form looks like  a canonical  one:
\eqn\smfr{
\omega = \sum_{i=1}^{m} dp_{i} \wedge dq^{i}
}
The local canonical coordinates are defined
up to the symplectomorphisms. Unlike the general diffeomorphisms,
which have $N$ functional degrees of freedom, $N$ being the dimension
of the manifold, the symplectomorphisms have only $1$ functional
degree of freedom.

The evolution of a Hamiltonian system is defined with the help
of Hamiltonian $H: M \to \IR$. The  function $H$ defines
a Hamiltonian vector field by the formula
\eqn\hmvc{
\iota_{V_{H}} \omega = - dH
}
The integrable system  on $M$ has a maximal collection of the
functionally independent
commuting Hamiltonians $H_{i}$ $i = 1, \dots , m = {\half}{\rm dim}M$ :
\eqn\cmt{
[ V_{H_{i}}  , V_{H_{j}} ] = 0
}
Let ${\ih}: M \to B \approx \IR^{m}$ be the map defined as:
${\ih}: x \mapsto \left( H_{1} (x), \ldots,
H_{m}(x) \right)$.
Liouville's theorem states that the integrable system has
a normal form locally: there are
coordinates $(I_{i}, \varphi^{i})$, such that
\eqn\nrmfrm{\eqalign{&
\omega = \sum_{i} dI_{i} \wedge d\varphi^{i} \cr
&
H_{k} = f_{k} ( \{ I \}) \cr}}
i.e. $I_{k}$ are coordinates on $B$.
For a sufficiently small domain  $U \subset B$
the space ${\ih}^{-1}(U)$ is the product $U \times \IR^{n-m} \times
T^{2m - n}$ and $\varphi^{k}$ are standard linear coordinates on
$\IR^{n-m} \times
T^{2m - n}$.
 If   the common level set of all Hamiltonians
is compact then this set is isomorphic to the torus
of the dimension $m$. In that case one may impose a condition on
the coordinates $\varphi^{i}$
that the differentials $d\varphi^{i}$
have periods which are integer multiples of $2\pi$. This fixes the
coordinates $(I, \varphi)$ up to the action of discrete group
${\rm PGL}_{m}({\IZ})$.

The Liouville theory also has a counterpart in the holomorphic setting
where the manifold $M$ is replaced by the complex manifold,
the symplectic form is a holomorphic
closed $(2,0)$-form, the Hamiltonians $H$ are
the holomorphic functions and the vector fields are
holomorphic vector fields. The Liouville theorem modifies
in this case. In fact the Liouville real tori are
replaced by the complex tori. If we require these tori to be abelian
varieties then we get what is called algebraically integrable system
\donagirev.
In the family
of such varieties the degenerate fibers can
appear.

The coordinates $I_{i}$ are referred to as ``action'' variables.
If $n=m$ then there is a nice formula for $I$.
Let $b_{1}, b_{2} \in U \subset B$ be sufficiently close to each other.
Choose a basis $e_{b}$ in ${\IH}_{1}({\ih}^{-1}(b_{1}), \IZ)$.
Connect
the points $b_{1}$ and $b_{2}$  with a path $\gamma \subset U$. The base $e_{b_{1}}$
can be transported to ${\IH}_{1}( {\ih}^{-1}(\gamma), \IZ)$ by
means of the Gau{\ss}-Manin
connection and it defines an element $\vec \Gamma \in
{\IH}_{2}({\ih}^{-1}(\gamma),
{\ih}^{-1}(b_{1} \cup b_{2}); \IZ)$. Then
$$
\vec I ( b_{2}) - \vec I (b_{1}) = \int_{\vec  \Gamma} \omega
$$

\subsec{ $(p,q) \to (I, \varphi)$}

Suppose that we have {\it  two integrable systems $\{H_k\}$ and $\{H^D_k\}$ on
the same symplectic manifold $M$. In this situation we say that these two
systems are dual to each other.}

Notice that this definition does not make duality an involution.

A pair of integrable systems given on one symplectic manifold $(M,\omega)$
is called {\it self-dual} if there exists a symplectic involution
$\sigma: M\rightarrow M$ exchanging $\{H_k\}$ and $\{H^D_k\}$, i.e., such
that for any $k = 1, \ldots, n$
$$ \sigma^* H_k = H_k^D $$

Once we have two integrable systems on the same manifold such that
both collections of Hamiltonians constitute at least locally a coordinate
system on the phase space $M$, one can write down the equations of motion
of  the second integrable system in the second order formalism using
the action variables $I_{i}$ of the first system as the coordinates $q^{i}$
for the second.

The global version of this definition is:
{\it Two Hamiltonian systems are dual to each other  in the sense of
{\sl action-coordinate} (AC) duality if the action variables $I_{i}$ of the first
system coincide with the coordinates $q^{i}$ of the second one and vice versa}.

\subsec{ $I \to I^{D}$}

In the holomorphic algebraic category there is an interesting complication:
the torus has a complex dimension $m$ and therefore
${\IH}_{1}({\bf T}_{b}; {\IZ}) = {\IZ}^{2m}$.

Again, let $b_{1}, b_{2} \in U \subset B$ be sufficiently close to each other.
Choose a symplectic basis
$e_{b_1} = (A_{\alpha}, B^{\beta})$ in ${\IH}_1 ({\ih}^{-1}(b_1), {\IZ})$
such that
$ A \cap A = B \cap B = 0, A_{\alpha}
\cap B^{\beta} = \delta^{\beta}_{\alpha}$, where $\cap$
is an intersection form ${\IH}_{1} \otimes_{\IZ} {\IH}_{1} \to {\IZ}$.
Connect
the points $b_{1}$ and $b_{2}$  with a path $\gamma \subset U$. The base $e_{b_{1}}$
can be transported to ${\IH}_{1}( {\ih}^{-1}(\gamma), \IZ)$ by
means of the Gau{\ss}-Manin
connection and it defines an element $\vec \Gamma \in
{\IH}_{2}({\ih}^{-1}(\gamma),
{\ih}^{-1}(b_{1} \cup b_{2}); \IZ)$.
The $(A,B)$-splitting of the basis $e_{b_1}$ is also transported
via Gau{\ss}-Manin along ${\gamma}$ and defines the decomposition
${\vec\Gamma} = {\vec\Gamma}_{A} + {\vec\Gamma}_{B}, \quad
{\vec\Gamma}_{A,B} \in {\IH}_{2}({\ih}^{-1}(\gamma),
{\ih}^{-1}(b_{1} \cup b_{2}); \IZ)$. Then define
\eqn\abprds{\eqalign{&
{\vec I} ({b_{2}}) - {\vec I} ({b_{1}}) = \int_{\vec\Gamma_{A}} \omega \cr
& {\vec I}^{D} ({b_{2}}) - {\vec I}^{D} ({b_{1}}) = \int_{\vec\Gamma_{B}} \omega \cr}}
We can take as the action variables the components of $\vec I$.
The reason for the $B$-cycles to be
discarded is simply the fact that the $B$-periods $\vec I^{D}$
are not independent of the $A$-periods. On the other
hand, one can choose as the independent periods the
integrals of $\omega$ over any Lagrangian
(in the sense of $\cap$) sublattice in ${\IH}_{1}({\bf T}_{b}; {\IZ})$
transported along $\gamma$.

This leads to the following structure of the action variables
in the holomorphic setting. Locally over a patch  in $\bf B$ one
chooses a basis in ${\IH}_{1}$ of the fiber together with the
set of $A$-cycles. This choice may differ over another
patch. Over the intersection of these discs one has
a ${\rm Sp}_{2m}({\IZ})$ transformation relating the bases. Altogether
they form an ${\rm Sp}_{2m}({\IZ})$ bundle. It is an easy excercise on the
properties of the period matrix of abelian varieties that the two form:
\eqn\tfrm{dI^{i} \wedge dI^{D}_{i}}
vanishes. Therefore one can always locally  find a function
$\CF$ - {\it prepotential}, such that:
\eqn\prptnt{I^{D}_{i} = {{\p \CF}\over{\p I^{i}}}}

{\it This duality maps the integrable system to itself.
It is called {\sl action-action} (AA) duality.}

\subsec{Quantum duality}

There exists a clear quantum counterpart of this picture.
Consider the eigenvalue problem for the Schr\"odinger
operators and the issue of the normalization of the wave-functions.

The quantum integrable system is a complete
collection of  ``independent'' (in the appropriate
sense) commuting
operators $\{ {\hat H}_{i} \}, i = 1, \ldots, m$, acting in the
Hilbert space $\CH$ of the model. By completeness we
mean that these operators have simple common spectrum $U \subset {{\IR}}^m$.The
commuting operators have common eigenfunctions. Generically
the eigen-value problem:
\eqn\egpr{
{\hat H}_{k} \vert {\vec \lambda} \rangle = e_{k} ({\lambda}) \vert {\vec
\lambda} \rangle }
has the unique (up to normalization) solution. Here $e_{i}$ is the corresponding
eigenvalue and ${\vec \lambda}$ is a label, which takes values in some  set $\Lambda$.
Altogether, $e_{k}$ form an imbedding  $e_{*}: \Lambda \to U$

Typically one has another set of commuting operators
(``position operators'') $\{ {\hat H}_{k}^{D} \}$ and the eigen-states
$\vert {\vec x} \rangle$ with eigen-values $e_{k}({\vec x})$
in the Hilbert   space of the model are represented as
the appropriate functionals on the space of the eigen-values
of the  operators $\hat H_{k}^{D}$. Here $\vec x$ is another label, which takes
values in the set $\Lambda^{D}$, which is mapped by $e^{D}_{*}$ to $U^{D} \in {\IR}^{m}$.
The familiar
case is $M = T^{*}{\CM}$, $\CH_{\CM} = L^{2}({\CM})$,
 the operators $\hat H_{i}$ are represented as
commuting
differential  operators, and ${\hat H}^{D}_{k}$ are represented as operators of multiplication
by a function $e^{D}_{i}(\vec x)$.
Suppose that  we are given  two classically $AC$ dual
Hamiltonian
systems. Let $I_{i}$ and $I^{D}_{i}$ be
their action variables (recall that $I^{D}_{i}$ are the coordinates for the first system).
Assume that there exist quantum integrals of motion
for both systems. Let us denote by  $\hat I_{i}$  the quantum
integrals of motion of the first system and by
$\hat I_{i}^{D}$  the  quantum integrals
of motion of the second one.

Once we have a quantum integrable system we can identify ${\CH}$
with the space $L^2(\Lambda)$ of square integrable functions on $\Lambda$ (w.r.t. the
spectral measure $d\mu$).  Indeed, choose a basis in $\CH$ consisting of
the common eigenvectors \egpr\ $\vert \vec{\lambda}\rangle$, where $\vec{\lambda}
= \{ \lambda_1, \ldots, \lambda_m \} \in \Lambda$. Then for any $|\psi\rangle \in {\CH}$
one can associate the function $\langle\vec{\lambda}| \psi\rangle$ on $\Lambda$.
Of course this
mapping ${\CH} \rightarrow L^2 (\Lambda)$ depends on the normalization of the
eigenvectors we have chosen.

In particular any operator $A$ acting on $\CH$ can be expressed as an
operator acting on $L^2(\Lambda)$ as
$$\hat{A}:\psi(\vec{\lambda}) \mapsto
\int_{\Lambda}\langle\vec{\lambda}|A|\vec{\lambda}^\prime\rangle
\psi(\vec{\lambda}^\prime)d\mu(\vec{\lambda}^\prime)
$$
\noindent
Now suppose that we have two integrable systems $\hat H_1, \ldots, {\hat H}_{n}$ and
$\hat H^D_1, \ldots, {\hat H}^{D}_{n}$ on the same Hilbert space $\CH$. We can use the first
one to identify the Hilbert space with the space of functions on its spectrum
$L^2(\Lambda)$ and write down the operators of the second integrable system as
acting on these functions and not on some abstract Hilbert space vectors.
Consider the function $\langle\vec{\lambda}|\vec{\lambda}^D\rangle \in
L^2(\Lambda) \otimes L^2(\Lambda^D) = L^2(\Lambda \times \Lambda^D)$, where
$|\vec{\lambda}\rangle$ and $|\vec{\lambda^D}\rangle$ are the eigenvectors of the
first and the second integrable system respectively. This function by
definition satisfies for any $k =1,\ldots, m$ the following equations:
\eqn\eitwo{\eqalign{&
\int_{\Lambda}\langle\vec{\lambda}|\hat H_i^D|\vec{\lambda}^{\prime}\rangle
\langle\vec{\lambda}^{\prime}|\vec{\lambda}^D\rangle d\mu({\vec \lambda^{\prime}}) =
e_{i}^{D}(\lambda^D) \langle\vec{\lambda}|\vec{\lambda}^D\rangle\cr
& \int_{\Lambda^{D}}\langle\vec{\lambda}|\vec{\lambda}^{D\prime}\rangle
\langle\vec{\lambda}^{D\prime}|\hat H_i|\vec{\lambda}^D\rangle d\mu({\lambda^{D\prime}}) =
e_{i}(\lambda)\langle\vec{\lambda}|\vec{\lambda}^D\rangle \cr}}
We see that the function $\langle\vec{\lambda}|\vec{\lambda}^D\rangle$ turns out to
be an eigenfunction for two commuting set of operators acting on two groups
of variables.  Otherwise
the function $\langle\vec{\lambda}|\vec{\lambda}^D\rangle$ is not uniquely defined by the two
dual integrable systems since the arbitrary change of the normalizations of
the eigenvectors $|\vec{\lambda}\rangle \mapsto
F(\vec{\lambda})|\vec{\lambda}\rangle$ and $|\vec{\lambda}^{D}\rangle \mapsto
F^D(\vec{\lambda}^D)|\vec{\lambda}^D\rangle$
results in $$ \langle\vec{\lambda}|\vec{\lambda}^D\rangle \mapsto
\overline{F(\vec{\lambda})}F^D(\vec{\lambda}^D)
\langle\vec{\lambda}|\vec{\lambda}^D\rangle
$$

Note also that though we have written the equations \eitwo\
as integral equation with the kernel being a generalized function for
particular dual systems these equations may be differential or difference
ones. (In fact it happens in  examples considered in the sequel.)

Analogously to the classical case a pair of quantum integrable systems
$\{H_i\}$ and $\{H_i^D\}$ is called self-dual if there exists a unitary
involution $\sigma:{\CH}\rightarrow {\CH}$
exchanging the two collections of operators, i.e. such that for any $k =
1,\ldots,m$ $$\sigma \cdot H_k = H_k^D\cdot  \sigma $$
Though the two integrable systems of a self-dual pair are completely
equivalent, the corresponding function $\langle\vec{\lambda}|\vec{\lambda}^D\rangle$
does not necessarily satisfy the condition
$\langle\vec{\lambda}|\vec{\lambda}^{D}\rangle=
\overline{\langle\vec{\lambda}^D|\vec{\lambda}\rangle}$. One can make it obey this
equation after a suitable normalization of the eigenvector bases.

In some cases it is natural to choose $\lambda$ among the action variables of the classical integrable system.

%% file: dual3.tex
%%%%%%%%%%%%%%%%%%%%
%%%%%Glava 3 %%%%%%%%%%%
%%%%%%%%%%%%%%%%%%%%%
%%% Dvukhchasticnye sistemy %%%%%
%%%%%%%%%%%%%%%%%%%%%%
%%%%%%%%%%%%%%%%%%%%%%
\newsec{Examples of dual systems. One degree of freedom:}

In this section we work out explicitly a few examples of the dual systems.

\subsec{Classical systems}

Two-particle systems which we are going to consider  reduce (after
exclusion of the center of mass motion) to a one-dimensional problem.
The action-angle variables can  be written explicitly  and
the dual system emerges immediately once the natural Hamiltonians
are chosen. The problem is the following. Suppose
the phase space is coordinatized by $(p,q)$. The dual Hamiltonian
(in the sense of AC duality) is a function of $q$
expressed in terms of $I, \varphi$, where $I, \varphi$ are the action-angle
variables of the original system :  $H_{D}(I, \varphi) = H_{D}(q)$. In all
the cases  below there is
a natural choice of $H_{D}(q)$.

\noindent
\sssec{Calogero \quad oscillator.} The
Hamiltonian in the center of mass frame reads as:
\eqn\clms{
H (p,q)  = {{p^{2}}\over{2}} +
{{\omega^{2} q^{2}}\over{2}} +  {{{\nu}^{2}}\over{2q^{2}}}
}
where $\omega$ and $\nu$ are the parameters. In the limiting
cases $\nu = 0$ and $\omega = 0$ one gets the usual
oscillator and rational Calogero-Moser system respectively.
The action-angle variables $I, \varphi$ can be found
by the standard procedure:
\eqn\aclmi{\eqalign{
I = {1\over{2\pi}} \oint pdq =  {1\over{2\pi}} \oint & \sqrt{ 2(E -
{{\omega^{2} q^{2}}\over{2}} -  {{{\nu}^{2}}\over{2q^{2}}})} dq \cr
d\varphi = & {{dq}\over{p}} \Bigl( {{\p I}\over{\p E}}\Bigr)^{-1}_{I} \cr}}
with the result:
\eqn\aclmii{\eqalign{
I = {{E - \omega\nu}\over{2\omega}} = &{1\over{4\omega}}
\biggl[ p^{2} + (\omega q - {{\nu}\over{q}})^{2} \biggr]\cr
H_{D}(I, \varphi) = {{q^{2}}\over{2}} = &{{I}\over{\omega}} \biggl[
1 + {{\nu}\over{2I}} + \sqrt{1 + {{\nu}\over{I}}} \cos{\varphi} \biggr]\cr}
}
The limit
$\nu \to 0$ is straightforward, yet tricky.
We must rescale $\varphi \to 2\varphi$ since the period of motion
jumps as $\nu$ approaches zero. We get:
\eqn\oscd{\eqalign{& I = {{E}\over{2\omega}} \cr
& H_{D}(I,\varphi) = q = 2\sqrt{I} \cos({\varphi})\cr}}
The limit $\omega \to 0$ is more
subtle as the classical motion becomes infinite.
For the system with the Hamiltonian
\eqn\rcm{
H(p,q) = {{p^{2}}\over{2}} + {{\nu^{2}}\over{2q^{2}}}
}
the action variable could be defined as
the asymptotic value of  the momentum:
$I = \sqrt{2E}$. This choice gives rise to the evolution, linear in the
``angle''-like variable,
\eqn\ang{\eqalign{
\varphi = &\sqrt{q^{2} - {{\nu^{2}}\over{2E}}}\cr
H_{D}(I, \varphi) =& {{q^{2}}\over{2}} = {{\varphi^{2}}\over{2}} + {{\nu^{2}}\over{2I^{2}}}\cr}
}

\noindent
\sssec{Sutherland \quad model.} The Hamiltonian is:
\eqn\twosu{
H(p,q)  = {{\half}p^{2}}  + {{\nu^{2}}\over{2{\sin}^{2}(q)}}.}
The action variable $I$ can be chosen to be:
\eqn\twosuac{
I = \sqrt{2E} }
To  prove that one might go to the coordinate $t = \cos(q)$ and compute the
integral ${1\over{2\pi}} \oint pdq$ by residues.
The angle variable $\varphi$ can be  determined from the
condition $dp \wedge dq = d{\tilde I} \wedge d\varphi$.
We get:
\eqn\twosua{d\varphi ={{{I} dq}\over{\sqrt{ {I}^{2} -
{{2\nu^{2}}\over{{\sin}^{2}(q)}}}}} }
\eqn\twosus{
H_{D}(I, \varphi) = {\cos} (q) =
{\cos}{\varphi}
\sqrt{ 1 - {{2\nu^{2}}\over{{I}^{2}}}} }
Notice, that \twosus\  coincides with the
Hamiltonian of the rational Ruijsenaars model (see below).

\noindent
\sssec{Elliptic \quad Calogero-Moser \quad system.} The Hamiltonian is:
\eqn\twoel{
H(p,q) = {{p^{2}}\over{2}}  + {\nu^{2}} {\wwp}
}
Here $p,q$ are complex, $\wwp$ is the Weierstrass function on the elliptic curve $E_{\tau}$:
\eqn\dfwwp{
{\wwp} = {1\over{q^{2}}} + \sum_{\matrix{&(m,n) \in {{\IZ}^{2}} \cr  &(m,n) \neq  (0,0) \cr}}
{1\over{(q + m\pi  +n \tau\pi )^{2}}} - {1\over{(m\pi  +n \tau\pi)^{
2}}}
}
Let us introduce the Weierstrass notations: $x = {\wwp}$, $y = {\wwp}^{\prime}$.
We have an equation defining the curve $E_{\tau}$:
\eqn\crve{
y^{2} = 4x^{3} - g_{2}(\tau) x -  g_{3}(\tau) = 4 \prod_{i=1}^{3} (x - e_{i}), \quad \sum_{i=1}^{3} e_{i} = 0}
The holomorphic differential $dq$ on $E_{\tau}$ equals $dq = dx/y$.
Introduce the variable $e_{0} = E/{\nu}^{2}$.
The action variable is one of the periods of the
differential ${pdq}\over{2\pi}$ on the curve
$E = H(p,q)$ :
\eqn\twoelac{
I = {1\over{2\pi}} \oint_{A} dq  \sqrt{2(E - {\nu}^{2} {\wwp}) } =
{i{\nu}\over{2\sqrt{2}\pi}} \oint_{A} {{dx \sqrt{x - e_{0}}}\over{\sqrt{(x-e_{1})(x-e_{2})(x-e_{3})}}}
}
The angle variable can be determined from the condition $dp \wedge dq = dI \wedge d\varphi$:
\eqn\twoelan{
d\varphi = {1\over{2i T(E)}} {{dx}\over{\sqrt{\prod_{i=0}^{3}(x - e_{i})}}}
}
where $T(E)$ normalizes $d\varphi$ in such a way that
the  $A$ period of $d\varphi$ is
equal to $2\pi$:
\eqn\wer{T(E) ={1\over{4\pi i }}  \oint_{A} {{dx}\over{\sqrt{\prod_{i=0}^{3} (x - e_{i})}}}}
Thus:
\eqn\www{\eqalign{
2i T(E)
d\varphi \qquad &= \qquad {{dx}\over{\sqrt{4 \prod_{i=0}^{3}
(x - e_{i})}}}\cr
\omega d\varphi \quad &= \quad  {{dt}\over{\sqrt{4 \prod_{i=1}^{3} ( t- t_{i})}}}\cr}}
where
\eqn\pps{\eqalign{
\omega = - 2i T(E) \sqrt{e_{01}e_{02}e_{03}}   =&  {1\over{2\pi}} \oint_{A}  {{dt}\over{\sqrt{4 \prod_{i=1}^{3} ( t- t_{i})}}}\cr
t=  {1\over{x-e_{0}}} + {1\over{3}} \sum_{i=1}^{3} {1\over{e_{0i}}} \quad & t_{i} = {1\over{3}}
\sum_{j=1}^{3} {{e_{ji}}\over{e_{0i} e_{0j}}}\cr
& e_{ij} =  e_{i} - e_{j}\cr}}

Introduce a meromorphic function on $E_{\tau}$:
\eqn\cosel{
{\widehat {cn}}_{\tau}(z) = \sqrt{{x-e_{1}}\over{x-e_{3}}} }
where $z$ has periods $2\pi$ and $2\pi \tau$.
It is an elliptic analogue of the cosine (in fact, up to a rescaling
of $z$ it coincides with the Jacobi elliptic cosine).
Then we have:
\eqn\twoeld{
H_{D}(I, \varphi) = {\widehat {cn}}_{\tau}(z) = {\widehat {cn}}_{\tau_{E}}({\varphi})
\sqrt{1 -{{{\nu}^{2}e_{13}}\over{2E - {\nu}^{2}e_{3}}}}
}
where $\tau_
{E}$ is the modular parameter of the relevant spectral curve
$v^{2} = 4 \prod_{i=1}^{3} ( t - t_{i} )$:
\eqn\tudl{
\tau_{E} = \Bigl( \oint_{B} {{dt}\over{\sqrt{4 \prod_{i=1}^{3}
(t - t_{i})}}}\Bigr) \Bigr/ \Bigl( \oint_{A} {{dt}\over{\sqrt{4 \prod_{i=1}^{3}
(t - t_{i})}}} \Bigr).} For large $I$, $2E(I) \sim I^{2}$

\noindent
\sssec{Elliptic \quad Ruijsenaars \quad model.} The Hamiltonian is:
\eqn\hamelrs{H (p,q) = \cos({\beta} p) \sqrt{1- {2(\beta{\nu})^{2}}{\wp}_{\tau}(q) }.}
As the curve $E_{\tau}$ degenerates one flows down to the trigonometric
(${\wwp} \rightarrow {1\over{\sin^{2}(q)}}$) or rational ($\wwp \to
{1\over{q^{2}}}$) Ruijsenaars system.
The spectral curve $H(p,q)=  E$ helps to define  the action variable
$I$:
\eqn\actelrs{I = {1\over{2\pi}} \oint_{A} p dq,}
up to the transformations
$$I \to n_1 I^{D} + n_2 I + {{2\pi}\over{\beta}} n_3 $$
where $n_1, n_2, n_3 \in \IZ$ and $(n_1, n_2) = 1$.  These transformations
reflect the freedom in the choice of the cycle $A$ on the spectral curve.
The appearence of $n_3$ was used in \fivedim.
We can write an explicit formula for the quantity which is better defined:
\eqn\drvt{{{\p I}\over{\p E}} = {1\over{2{\sqrt{2}}\pi\beta^{2}\nu}}
\oint_{A} {{dx}\over{\sqrt{\prod_{i=0}^{3}(x - e_{i})}}} }
where now $e_{0} = {{1-E^{2} }\over{2(\beta{\nu})^{2}}}$. Under $A \to B$
transformation ${{\p I}\over{\p E}}$ gets multiplied
by $\tau_{E}$, where $\tau_{E}$ is defined as in \tudl\ .
Quite similarly to \www\ we get:
\eqn\vph{d\varphi = {1\over{T(E)}} {{dx}\over{\sqrt{\prod_{i=0}^{3}(x-e_{i})}}} }
with
\eqn\nrmlz{T(E) = {1\over{2\pi}} \oint_{A} {{dx}\over{\sqrt{\prod_{i=0}^{3}(x-e_{i})}}}}
Finally, for $H_{D}$ given by \cosel\ we get:
\eqn\dhmlsers{H_{D}(I, \varphi) = {\widehat {cn}}_{\tau}(z) = {\widehat {cn}}_{\tau_{E}}({\varphi})
\sqrt{1 - {{2{(\beta\nu)}^{2}e_{13}}\over{E(I)^{2}- 1 - 2{(\beta\nu)}^{2}e_{3}}} }
}
Asymptotically, for large $I$, $E(I) \sim \cos( \beta I)$.

\noindent
\sssec{General \quad elliptic \quad model.} In the general case one modifies
the formula \crve\ in such a way that the coefficients
$g_{2}$ and $g_{3}$ are the sections of the line bundles $\CO(4n)$ and
$\CO(6n)$ respectively over ${\IB} \approx {\IP}^{1}$.
The elliptic curve $E_{z}$ defined by
the modified \crve\ degenerates over the divisor of zeroes  of
its discriminant:
\eqn\dscrm{
\Delta = g_{2}^{3} - 27 g_{3}^{2}}
which is a section of $\CO(12n)$.
The latter has generically $12n$ zeroes. To make the total space of
fibration isomorphic to the $K3$ surface (compact simply-connected
symplectic surface)
we need $24$ singular fibers, i.e. $n=2$. The Hamiltonian of the
integrable system we consider is any function on $\IB$.
It gives rise to a meromorphic
vector field on $M$, which linearizes along the elliptic fibers.
The symplectic form is given by:
\eqn\symp{\omega =  {{dx \wedge dz}\over{y}}}
where $z$ is the projective coordinate on $\IB$.
Under change of  the variables:
$\tilde z = {1\over z}$, $\tilde y = -{y\over z^{6}}$, $\tilde x =
{x\over z^{4}}$
the form \symp\ goes over to ${{d\tilde x \wedge d\tilde z}\over{\tilde y}}$
and the equation
\crve\ is mapped to
$$
{\tilde y}^{2} = 4{\tilde x}^{3} - {\tilde g}_{2}({\tilde z}) {\tilde x} - {\tilde g}_{3}({\tilde z})
$$
where the polynomials ${\tilde g}_{k}$ are defined through the relation:
$$
{\tilde g}_{k} ({\tilde z}) = {\tilde z}^{4k} g_{k} (1/{\tilde z}).
$$

Over
a simply
connected region $U \subset {\IP}^{1} \backslash \Delta^{-1}(0)$
one can trivialize the bundle of the first homologies
${\IH}_{1}(E_{z}, {\IZ})$, in particular to make a well-defined choice of
the $A$-cycle of the elliptic fibre.
The local
action variable $I = I(z)$ is defined over $U$
by the equation:
\eqn\actnkthr{{{dI(z)}\over{dz}} = T(z) = {1 \over{2\pi}}
\oint_{A} {{dx}\over{y}}  }
where the integral is taken over a
chosen $A$-cycle.
The fibration of ${\IH}_{1}(E_{z}, {\IZ})$ over
$\IB \backslash \Delta^{-1}(0)$ is non-trivial and there is no global
monodromy invariant choice of $A$-cycles. So the action variable is defined
by \actnkthr\ only locally. The monodromies around the degenerate fibers
corresponding to various singularities has been worked out by Kodaira and their physical
interpretation can be found in \GaMorSe.
For generic polynomials $g_{2}(z), g_{3}(z)$
the singularities are of the type $A_{1}$.

The angle variable
dual to $I(z)$
is nothing but the linear coordinate on
the Jacobian of the
fiber elliptic curve \crve. In particular
it is periodic with the periods
$2\pi $ and $2\pi\tau(z)$.
It  is to be found from the
relation:
\eqn\anglfrst{d\varphi = {1\over{T(z)}} {{dx}\over{y}}. }

We can
get a dual system by treating $x$ as the Hamiltonian.
Since $x$ is not a
meromorphic  function on $K3$ (it changes
under the $z \to {1\over z}$
transformation) this is
only possible if we delete the elliptic fiber
$E_{\infty}$.

Let us see what will
be the action-angle variables.
First
of all, generically the fiber $C_{x}$ over $x \in \IP^{1}$
is an incomplete
hyperelliptic curve of genus $5$.
The holomorphic differentials on this
curve are:
$$
\omega_{k} = {{ z^{k} dz}\over{\sqrt{4x^{3}- g_{2}(z) x -
g_{3}(z)}}}, \quad k = 0, \ldots 4
$$
The action variable $I^{D} = I^{D}(x)$
obeys the equation:
\eqn\actndl{{{dI^{D}}\over{dx}} = T^{D}(x) =
{1\over{2\pi}} \oint_{L}
 {{dz}\over{y}} }
where $L$ is a one-cycle in
${\IH}_{1}(C_{x}, {\IZ})$.
Here we face $AA$ duality in its
extreme form: the
freedom to choose $L$ is much bigger
then in the case of original
system,
since the corresponding duality group is ${\rm Sp}_{10}({\IZ})$. We can
partially
integrate \actndl\ to get
\eqn\actnint{I^{D} = {1\over{2\pi}}
\oint_{L} \varphi T(z) dz}  where
$$
x = {1\over{T(z)^{2}}} {\wp}
\left(\varphi ; \tau(z) \right)
$$
The angle variable is one of the linear
coordinates on the Jacobian variety of
$C_{x}$, which is $5$ dimensional
abelian variety:
$$
d\varphi^{D} = {1\over{T^{D}(x)}} {{dz}\over{y}}
$$
The
embedding of the Liouville tori into the abelian varieties of
higher rank
originating from hyperelliptic curves
is a well-known phenomenon in the
theory of integrable systems,
going back to the original work of S.~Novikov
and A.~Veselov \novves.

\subsec{Quantum systems.}

Here we work out a few  examples of
quantum dual systems.

\noindent
\sssec{Harmonic\quad oscillator.} The Hamiltonian
\clms\ in the limit $\nu = 0$ quantizes to:
\eqn\qosc{
{\hat H} = {-\half}{{{\p}^{2}}\over{{\p} q^{2}}} +
{{{\omega}^{2} q^{2}}\over{2}}}
Its
normalized eigen-functions are \landau :
\eqn\her{\eqalign{ {\hat H}
{\psi}_{n}  = & {\omega} (n + {\half}) \psi_{n}\cr
{\psi}_{n} (q) =  \bigl(
{{\omega}\over{\pi}} \bigr)^{1/4}  & {{
e^{-{{\omega
q^{2}}\over{2}}}
}\over{2^{n/2}\sqrt{n!}}} H_{n}(q \sqrt{\omega}) \cr}}
where
$H_{n}(\xi)$ is the Hermite polynomial:
$H_{n}(\xi) = e^{\xi^{2}} (-
\p_{\xi})^{n} e^{-{\xi}^{2}}$.
Using this representation of the
wave-function one can easily obtain a reccurence
relation (details are in the
appendix):
\eqn\recrel{
\sqrt{n+1} {\psi}_{n+1}(q) + \sqrt{n} {\psi}_{n-1}
(x) = \sqrt{2\omega} q {\psi}_{n}(x)}
It means that $\psi_{n}(q)$ is an
eigen-function of the following
difference operator:
\eqn\dddop{
{\hat H}_{D}
= T_{+} \sqrt{n} + \sqrt{n} T_{-}, \quad T_{\pm} = e^{\pm {{\p}\over{\p
n}}}
}acting on the  subscript $n$. It is easy to recognize in \dddop\ the
quantized
version of \oscd.

\noindent
\sssec{Sutherland \quad model.} Here we deal
with the Hamiltonian:
\eqn\qsuha{
{\hat H} = {-\half }{{{\p}^{2}}\over{{\p}
q^{2}}} + {{{\nu}({\nu} -1)}\over{2 \sin^{2}(q)}}
}
Its normalized
eigen-functions are \landau:
\eqn\wwwsu{\eqalign{{\hat H} \psi_{n}  =
{{n^{2}}\over{2}} \psi_{n}&  \cr
\psi_{n}(q) = \sin^{\nu}(q) &  \sqrt{n
{{(n-{\nu})!}\over{(n+{\nu}-1)!}}} \Pi_{n-\half}^{\nu -
\half}\bigl(\cos(q)\bigr)\cr
& \Pi_{l}^{m}(x) = {1\over{l!}} {\p}_{x}^{l+m}
\bigl({{x^{2}-1}\over{2}}\bigr)^{l}\cr} }
For simplicity we take $\nu$ and
$n$ to be half-integers.
One can change $\nu \to - \nu - 1$ to get
another
eigen-function with the same eigenvalue.
Using the fact that the
generating function for $\Pi_{l}^{0}$'s is
\eqn\gnfn{
Z(y,x) =
\sum_{l=0}^{\infty} y^{l} \Pi_{l}^{0} = {1\over{\sqrt{1 - 2xy + y^2}}}}
one derives the
recurrence relations (details are in the appendix):
\eqn\retwoss{\eqalign{
&x
\Pi_{l}^{m} = {{l+1-m}\over{2l+1}} \Pi_{l+1}^{m} + {{l+m}\over{2l+1}}
\Pi_{l-1}^{m}\cr
\cos(q) \psi_{n} =& {\half} \Bigl( \sqrt{1-
{{\nu}({\nu}-1)}\over{n(n+1)}}\psi_{n+1}
+\sqrt{1-
{{\nu}({\nu}-1)}\over{n(n-1)}}\psi_{n-1} \Bigr)\cr}}
that is $\psi_{n}$ is an
eigen-function of the finite-difference operator
acting on the $n$
subscript:
\eqn\dddssop{\eqalign{&{\hat H}_{D} \psi(q) = \cos(q) \psi(q)\cr
{\hat H}_{D} &= T_{+} \sqrt{1- {{{\nu}({\nu}-1)}\over{n(n-1)}} }+
\sqrt{1- {{{\nu}({\nu}-1)}\over{n(n-1)}} }T_{-} \cr}} which is  a quantum
version of \twosus.

\noindent
\sssec{Moral \quad of \quad  the \quad story.}  The
moral of the previous discussion
is that the polynomial dependence on momenta
of the hamiltonian is traded with
the rational potential of the dual system.
The trigonometric potential is mapped
to the trigonometric (= relativistic)
dependence on momenta of dual system.
The elliptic potential gives rise to
elliptic (=``double-relativistic'' ) dependence
on momentum of the dual
system Hamiltonian. When the system with
trigonometric dependence on momentum
is quantized its Hamiltonian becomes
a finite- difference operator. The
wave-functions become the functions
of the discrete variables. The origin of
this is in the Bohr-Sommerfeld quantization
condition. Indeed, since the
trigonometric dependence of momenta
implies that the leaves of the
polarization are compact and moreover
non-simply connected the covariantly
constant
sections of the prequantization connection along the polarization
fiber generically seases
to exist. It is only for special ``quantized''
values of the action variables that the section exists.
In the elliptic case
the quantum
dual Hamiltonian is going to be a difference operator of
the infinite order.
The self-dual elliptic many-body system is still to be
constructed. It seems that to achieve this goal one needs a notion
of the Heisenberg double for the central extension of the
two dimensional current group \efren.

\noindent
\sssec{Example \quad of \quad the \quad prepotential.}
To illustrate the meaning of the {\sl AA} duality we look at the
 two-body system, relevant for the
$SU(2)$ $\CN=2$ supersymmetric
 gauge theory \gmmm:
\eqn\tdhml{H =
 {{p^{2}}\over{2}} + \Lambda^{2} \cos (q)}
with $\Lambda^{2}$ being a complex
 number - the coupling constant of a two-body
problem and at the same time a
 dynamically generated scale of the gauge theory.
The action variable is
 given by one of the periods of the differential $pdq$.
Let us introduce more
 notations:
$x = \cos (q)$, $y = {{p\sin (q)}\over{\sqrt{-2}\Lambda}}$, $u =
 {{H}\over{\Lambda^{2}}}$. Then the spectral curve, associated to the system
 \tdhml\ which is also a level set of the Hamiltonian
can be written as
 follows:
\eqn\spctrlcrve{y^{2} = (x - u) (x^{2}-1)}
which is exactly
 Seiberg-Witten curve \SeWi\ as it was first observed in \gmmm.
The periods
 are:
\eqn\actn{\eqalign{& I = \int_{-1}^{1} \sqrt{{x-u}\over{x^{2}-1}} {dx},
 \cr
& I^{D} = \int_{1}^{u} \sqrt{{x-u}\over{x^{2}-1}} {dx} \cr}}
They obey  Picard-Fuchs equation:
$$
\left( {{d^{2}}\over{du^{2}}} +
 {1\over{4(u^{2}-1)}} \right) \pmatrix{& I\cr& I^{D}\cr} = 0
$$
which can be
 used to write down an asymptotic expansion of the action
variable near
 $u=\infty$ or $u= \pm 1$ as well as that of prepotential \prptnt. The {\sl
 AA} duality is manifested in the fact that
near $u =\infty$ (which
 corresponds to the high energy scattering in the two-body problem
and also a
 perturbative regime of $SU(2)$ gauge theory) the appropriate action
 variable
is $I$ (it experiences a monodromy $I \to - I$ as $u$ goes around
 $\infty$), while
near $u = 1$ (which corresponds to the dynamics of the
 two-body
system near the top of the potential and to the strongly coupled
 $SU(2)$ gauge theory)
the appropriate variable is $I^{D}$ (which corresponds
 to a weakly coupled magnetic $U(1)$ gauge theory and is actually well
 defined near $u=1$ point) \SeWi. The monodromy
invariant combination of the
 periods \matone:
\eqn\mndr{II^{D} - 2\CF = u} (whose origin is in the periods of
 Calabi-Yau manifolds on the one hand
and in the  properties of anomaly in theory on the other) can be chosen
as a global coordinate on the space of integrals of motion $\bf B$.
At $u \to \infty$ the prepotential has an expansion of the form:
$$
\CF \sim {\half}  u \log u + \ldots \sim I^{2} \log I + \sum_{n}{{f_{n}}\over{n}} I^{2-4n}
$$
\subsec{Appendix.} To derive the recurrence relation for
the oscillator wave-functions we use the creation operator representation:
$\psi_{n} = {1\over{\sqrt{2n}}} ( -\p_{\xi} + {\xi}) \psi_{n}$.  Applying this relation
twice and using the fact that $\psi_{n}$ is an eigen-function
of $\hat H$ one arrives at  \recrel. For the Sutherland model  we use two obvious relations:
\eqn\obreli{(x-y)\p_{x} Z = y\p_{y} Z}
\eqn\obrelii{(1 - 2xy + y^{2})\p_{y} Z = (x-y)Z}
Next, \obreli\ implies:
\eqn\reli{
(y\p_{y} - m) \p_{x}^{m} Z = (x- y) \p^{m+1}_{x} Z}
and \obrelii\ yields:
\eqn\relii{ \bigl( (1 - 2xy + y^{2}) \p_{y}  + y - x \bigr) \p^{m}_{x}Z = m ( 1 + 2y\p_{y}) \p^{m-1}_{x} Z}
Combination of those two gives rise to \retwoss.
\vfill\eject
%%%% Konec glavy 3 %%%%%%%%

%% file: dual4.tex
%%%%%%% Glava 4 %%%%%%%
%%%%%%%%%%%%%%%%%%
%% mnogochastichnye sistemy %%%
%%%%%%%%%%%%%%%%%%%
\newsec{Duality in Many-Body Systems:}
In the previous sections we discussed the concepts of duality and
worked out explicitly several examples of dual  two-body systems in both  classical and quantum cases. We now turn to a study of
 many-body systems. The many-body systems can be divided into three
classes:
rational,
trigonometric and elliptic ones.
The Hamiltonian of the model
may depend on momenta/coordinates in any one of these three fashions.
The duality transformation  exchanges them.

\subsec{Examples.}

We summarize the systems
and their duals in the following table:
\eqn\table{\matrix{& &{\rm rat. CM} &
\leftrightarrow & {\rm rat. CM} & &\cr
& R \to 0 &   \uparrow  &
                 &  \uparrow       & \beta \to 0& \cr
& &{\rm trig. CM} & \leftrightarrow & {\rm rat. RS} & &\cr
& \beta \to 0 &   \uparrow  &
                        &  \uparrow       &R \to 0&\cr
& &{\rm trig. RS} & \leftrightarrow & {\rm trig. RS} & &\cr}}
Here $CM$ denotes {\it Calogero-Moser} models \c\m\s\ and
$RS$ stands for {\it Ruijsenaars-Schneider} \rs\r\raa\rrev\ruijdu.
The parameters $R$ and $\beta$ here are the radius of the circle
the coordinates of the particles take values in and the inverse speed
of light respectively.
The horizontal arrows in this table are the dualities, relating
the systems on the both sides. Most of them were discussed by
Simon Ruijsenaars
\ruijdu,\r. We
notice that the duality transformations form a
group which  in the case of self-dual systems listed here contains
${\rm SL}_{2}({\IZ})$.
The generator $S$ is the gorizontal arrow described below, while the
$T$ generator is in fact a certain finite time evolution of the
original system (which is always
a symplectomorphism, which maps
the integrable system to the dual one).
\noindent
We begin with recalling the Hamiltonians of these systems.
Throughout this section
$q_{ij}$ denotes $q_{i} - q_{j}$.

\noindent
\sssec{Rational \quad CM \quad model.} The phase space is
$(T^{*}V)/{\Gamma} $, where $V$ is a linear space acted
on by a Coxeter group $\Gamma$.
 We consider the
simplest case $V = {\IR}^{N}$, $\Gamma = \CS_{N+1}$.
Let $(p_{i}, q_{i})$ be the set of coordinates, $i = 1, \dots, N+1$
with the constraint $\sum q_{i} = \sum p_{i} = 0$.
The Hamiltonians can be conveniently
packaged using the Lax operator:
\eqn\lxrcm{\eqalign{H_{k} &={1\over{k}} {\Tr}L^{k}\cr
L_{ij} =& p_{i} \delta_{ij}  +
{{i{\nu}(1- \delta_{ij})}\over{q_{i}- q_{j}}} \cr}}
In particular, the quadratic Hamiltonian reads:
\eqn\qhmrcm{
H_{2} = \sum_{i} {\half} p_{i}^{2} + \sum_{i < j} {{{\nu}^{2}}\over{q_{ij}^{2}}}.}

\noindent
\sssec{Trigonometric \quad CM  = Sutherland \quad model}.
The phase space is
$(T^{*}V)/{\hat \Gamma}$, where $V$ is a linear space acted on by an affine
Coxeter group $\hat \Gamma$.
 We consider the
simplest case $V = {\IR}^{N-1}$, $\hat \Gamma = \CS_{N} \times {{2\pi}\over{R}}  {\IZ}^{N}$.
Let $(p_{i}, q_{i})$ be the set of coordinates, $i = 1, \dots, N$
with the constraint $\sum q_{i} = \sum p_{i} = 0$, and the identifications
 $q_{i} \sim
q_{i} + {{2\pi}\over{R}} n_{i}, n_{i} \in {\IZ}$.
The Hamiltonians can be conveniently
packaged using the Lax operator:
\eqn\lxtcm{\eqalign{H_{k} &={1\over{k}} {\Tr}L^{k}\cr
L_{ij} =&p_{i} \delta_{ij}  +
 {{iR{\nu}(1- \delta_{ij})}\over{2 \sin\bigl( {{R(q_{i}- q_{j})}\over{2}}\bigr)}} \cr}}
In particular, the quadratic Hamiltonian equals:
\eqn\qhmtcm{
H_{2} = \sum_{i} p_{i}^{2} + \sum_{i < j} {{R^{2}{\nu}^{2}}\over{4\sin^{2}\bigl( {{R(q_{i}- q_{j})}\over{2}}\bigr)}}.}

\noindent
\sssec{Rational \quad RS  = Relativistic \quad rational \quad CM \quad
model}.
The phase space is
$(T^{*}V)/{\hat \Gamma}$, where $V$ is a linear
space acted on by an affine
Coxeter group $\hat \Gamma$.
 We consider
the
simplest case $V = {\IR}^{N-1}$, $\hat \Gamma = \CS_{N} \times
{{2\pi}\over{\beta}}{\IZ}^{N}$.
Let $(p_{i}, q_{i})$ be the set of
coordinates, $i = 1, \dots, N$
with the constraint $\sum q_{i} = \sum p_{i} =
0$, and the identifications
 $p_{i} \sim
p_{i} + {{2\pi}\over{\beta}} n_{i},
n_{i} \in {\IZ}$.
The Ha miltonians can be conveniently
packaged using the
Lax operator:
\eqn\lxrrs{\eqalign{H_{k} &={1\over{k}} {\Tr}L^{k}\cr
 L_{ij} =
e^{- i\beta p_{i}} {{\beta\nu}\over{q_{ij} + {\beta\nu}}}& \prod_{k \neq j}
\sqrt{1 - {{{(\beta\nu)}^{2}}\over{q_{jk}^{2}}}}\cr}}
In particular, the
Hamiltonian ${\half} (H_{1} - H_{-1})$ equals:
\eqn\qhmrrs{
H ={\half}{\Tr}(L + L^{-1}) = \sum_{i} \cos ({\beta}p_{i})
\prod_{ j \neq i } \sqrt{ 1-{{{(\beta\nu)}^{2}}\over{q_{ij}^{2}}}}.
}
The Lax
operator \lxrrs\ is gauge equivalent to the operator
\eqn\lxrrstr{\eqalign{&
{\CL}_{ij} = e^{- i\beta p_{i}} {{\beta\nu}\over{q_{ij} +
{\beta\nu}}}
\sqrt{{\phi}^{+}_{i}{\phi}^{-}_{j}} \cr
& \phi_{i}^{\pm}  = \pm
{{{\Pi}(q_{i} \pm \beta\nu)}\over{\beta\nu {\Pi}^{\prime}(q_{i}) }},
\quad
{\Pi}(q) = \prod_{i} ( q - q_{i}) \cr}}
In the limit $\beta \to 0$ both
$L, \CL$ of \lxrrs,\lxrrstr\ behave as
$\Id - i\beta \,  ($~Lax operator in
\lxrcm~$) + o(\beta)$.

\noindent
\sssec{Trigonometric \quad RS = Relativistic \quad Sutherland \quad model.} The phase space
is
$(T^{*}V)/{\Gamma_{E}}$, where $V$ is a linear space acted on by a
double
affine
Coxeter group $\Gamma_{E}$, $E$ being an elliptic curve.
 We
consider the
simplest case $V = {\IR}^{N-1}$, $\Gamma = \CS_{N}
\times
\bigl({{2\pi}\over{\beta}} {\IZ}^{N} \oplus
{{2\pi}\over{R}}{\IZ}^{N}\bigr)$.
Let $(p_{i}, q_{i})$ be the set of
coordinates, $i = 1, \dots, N$
with the constraint $\sum q_{i} = \sum p_{i} =
0$, and the identifications
 $p_{i} \sim
p_{i} + {{2\pi}\over{\beta}} n_{i}, \quad
q_{i} \sim
q_{i} + {{2\pi}\over{R}} m_{i},\qquad n_{i}, m_{i} \in {\IZ}$.
The
Hamiltonians can be conveniently
packaged using the Lax
operator:
\eqn\lxtrs{\eqalign{H_{k} &={1\over{k}} {\Tr}L^{k}\cr
L_{ij} =
e^{-i{\beta} p_{i}}  {{\sin \left( {{R\beta\nu}\over{2}}
\right)}\over{\sin\bigl( {{R}\over{2}}( q_{ij} + {\beta\nu})\bigr)}}&
\prod_{k \neq j} \sqrt{1 - {{\sin^{2}\left({{R\beta\nu}\over{2}}\right)}\over{\sin^{2}\left({{Rq_{jk}}\over{2}} \right)}}}\cr}}
In particular, the  Hamiltonian ${\half} (H_{1} - H_{-1})$ equals:
\eqn\qhmtrs{
H ={\half}{\Tr}(L + L^{-1}) = \sum_{i} \cos ({\beta}p_{i}) \prod_{ j \neq i } \sqrt{ 1-{{\sin^{2}(R\beta\nu)}\over{\sin^{2}\bigl( {{R(q_{ij})}\over{2}} \bigr)}}}.
}
The Lax operator \lxrrs\ is gauge equivalent to the operator
\eqn\lxrtrstr{\eqalign{& {\CL}_{ij} = e^{- i\beta p_{i}} {{{\rm sin}\left( {{NR{\beta\nu}}\over{2}} \right)}\over{N {\rm sin}\left( {R\over{2}} (q_{ij} + {\beta\nu})\right)}}
\sqrt{{\Phi}^{+}_{i}{\Phi}^{-}_{j}} \cr
& \Phi_{i}^{\pm}  = \pm {{NR}\over{2{\rm sin}\left( {{NR\beta\nu}\over{2}} \right)}} {{{P}(q_{i} \pm \beta\nu)}\over{{P}^{\prime}(q_{i}) }}, \quad P(q) = \prod_{i=1}^{N} {\rm sin}
\left( {{R}\over{2}} (q -q_{i}) \right) \cr}}
In the limit $R \to 0$, with $\beta$ fixed the expressions
\lxtrs,\qhmtrs,\lxrtrstr\ naturally go over to \lxrrs, \qhmrrs, \lxrrstr\ respectively.
In the limit $\beta \to 0$, $R$
fixed both $L, \CL$ behave as $\Id - i\beta($~Lax operator
in \lxtcm~$) + o(\beta)$.

\subsec{ Explanations: Hamiltonian and/or
Poisson reduction}
Suppose we are given a symplectic manifold $(X, {\omega}_{X})$ with the
Hamiltonian action of a Lie group $G$ with equivariant moment map
${\mu}: X \to {\lieg}^{*}$. The symplectic quotient of $X$ with respect to
$G$ is the symplectic manifold $M$, denoted as $X//G$ and defined as:
$$
M = {\mu}^{-1}(0)/G
$$
Its symplectic form $\omega_{M}$ is defined through the relation:
$$
p^{*} {\omega}_{M} = i^{*} {\omega}_{X}
$$
where $p: {\mu}^{-1}(0) \to M$ is the projection and
$i: {\mu}^{-1}(0) \to X$ is the inclusion.

Let us assume that an integrable Hamiltonian system is defined
on $X$. Let ${\bar K} = \{ K_{1}, \dots , K_{x} \}$, $x = {1\over{2}} {\rm dim} X$
denote the set of its integrals of motion.
Suppose that this system is equivariant with respect to the action of $G$.
This is equivalent to the statement, that
$K_{i}$ and $\mu^{a}$ form a closed algebra $\CK$ with respect to
the Poisson brackets.
Let us assume that on the zero level of the moment map $\mu$
the center $Z({\CK})$ of  the algebra $\CK$ is sufficiently
big, i.e. the dimension of its spectrum equals  half  the  dimension
of $M$. Then the integrable system on
$X$ descends to the integrable system on $M$, $\CK$ being
replaced by $Z({\CK})$.

Now let us impose one further restriction. Suppose that
$X$ possesses another $G$-equivariant
integrable Hamiltonian system,
with integrals ${\bar Q} = \{ Q_{1}, \dots , Q_{x} \}$, which is dual
to the system ${\bar K}$ (algebraically
it means that $\bar K$ and $\bar Q$ generate
all functions on $X$). We also assume that ${\bar Q}$
descends to $M$.

On the original manifold $X$ the evolution of the system
$\bar K$ looks non-trivially in the action-angle variables for
the system $\bar Q$ and vice versa. The same is true
for the reduced systems.
The advantage of the consideration of $X$ is that the
systems on $X$ can be much simpler then those on $M$.
In the following sections we shall consider various
examples of this situation.

The similar statements hold in the case of Poisson manifolds, the
relevant reduction being the Poisson one (one first
takes a quotient with respect to the group
and then picks out a symplectic leaf). We leave the details to the
interested readers.

Now we proceed to the explicit constructions. We will discuss
 the models introduced in the previous section on case-by-case basis
and show how the reduction which yields these systems also explains
the dualities between different systems.

\noindent
\sssec{Rational \quad CM \quad model.}
This model can be obtained as a result of Hamiltonian reduction applied to
$T^{*}\lieg \times \CO$ \kks\
for $\lieg = su(N)$, $\CO = {\IC\IP}^{N-1}$. The symplectic form on this manifold
is the sum of  Liouville form on $T^{*}\lieg$ and $-N{\nu}\times$ Fubini-Study form on
$\CO$.
Let $(e_{1} :  \dots : e_{N})$
be the homogeneous coordinates on $\CO$. The group $G = SU(N)$ acts on $T^{*}\lieg$
via conjugation and on $\CO$ in a standard way ($\CO = G/H$, $H = S(U(N-1) \times U(1))$).
Then the moment map for the action of $\CG$
on
$T^{*}\lieg
\times \CO$ is
\eqn\mmprcm{\mu = ad_{Q}^{*}(P) - J \quad J_{ij} = {\nu} ( N \delta_{ij} - e_{i} e_{j}^{*}) }
where $Q \in \lieg, P \in \lieg^{*}$.
Now we choose two sets of Hamiltonians:
\eqn\dlhmrcm{
H_{k} = {1\over{k}} {\Tr} P^{k} \quad {\rm and} \quad H^{D}_{k} = {1\over{k}} {\Tr} Q^{k} }
If we identify $\lieg^{*}$ and $\lieg$ with the help of $\Tr$ then
the equation $\mu = 0$ has the form:
\eqn\mmrcmii{[P,Q] = J} which is obviously preserved by the  involution:
$P \rightarrow Q, Q \rightarrow - P$. So we are guaranteed to get a self-dual system. Now we
have to find suitable coordinates and action variables. Let us choose the
gauge (remember that we have to mod out \mmrcmii\ by the action of $G$):
\eqn\grcm{Q = {\rm diag}(q_{1}, \dots, q_{N})}
This gauge is preserved by the action of the maximal torus $T = U(1)^{N-1}$
which turns out to be sufficient to set all $e_{i}$ to be equal: $e_{i} = 1$ \perlbook.
Then the equation
\mmrcmii\ fixes $P$ which turns out to be nothing but $L$ in \lxrcm.
As it is obvious
that the reduced symplectic form equals $\sum_{i} dp_{i} \wedge dq_{i}$ (with the
constraint $\sum q_{i} = \sum p_{i} = 0$) one concludes that $q_{i}$'s are the action
variables for the system generated by $H_{k}^{D}$'s. Therefore eigenvalues of $P$ are the action
variables for the flows generated by $H_{k}$'s. We therefore proved the
following

\noindent
{\bf Statement.} {\it Consider the map:
\eqn\invkas{
{\sigma}: \{ (p_{i}, q_{i}) \} \to \{ (   \xi_{i} , - \eta_{i} )\}}
where $\eta_{i}$'s are the  eigenvalues of $L\equiv P$ and $\xi_{i}$ are the diagonal entries of
$Q$  in the eigenbasis of $P$. It is an involution.}

Let us go back to the systems \dlhmrcm. The moment map equation \mmrcmii\ is obviously preserved by the transformations of the form
\eqn\sltwol{(P, Q) \mapsto (aP + bQ, cP + dQ) \quad ad-bc=1}
which form $SL_{2}({\IR})$ group. The transformed Hamiltonians
$$
g\cdot H_{k} = {1\over{k}} {\Tr} (aP + bQ)^{k}
$$
are easy to express through the original Hamiltonians \dlhmrcm\ in the
coordinates $(p_{i}, q_{i})$:
$$
g\cdot H_{k} (p_{i}, q_{i}) \vert_{\nu} = H_{k} ( ap_{i} + bq_{i}, q_{i}) \vert_{a\nu}
$$
Let us restrict our attention to the ${\rm SL}_{2}({\IZ})$
subgroup of the group \sltwol.
It is generated by the transformations
\eqn\gnsltwo{S = \pmatrix{0 & -1\cr 1 &0\cr} \quad
T = \pmatrix{1 & 1\cr 0 & 1\cr}}
It is clear that $S$ coincides with the involution leading to
\invkas\ while
$T$ is the unit time  evolution with respect to the Hamiltonian $H_{2}^{D}$.

\noindent
\sssec{Trigonometric \quad CM , \quad Rational \quad RS}.
The
trigonometric $CM$ system can be obtained as
Hamiltonian reduction applied
to either $T^{*}G \times \CO$ \kks\ or
$T^{*}{\hat \lieg} \times \CO$ \gncal\
where ${\hat \lieg}$ is the central extension of the loop algebra.
In the
latter case one has to specify the action of the gauge group $LG$ on the
orbit
$\CO$. The correct choice is the most natural one: since the orbit is
finite-dimensional,
the only sensible way the loop group can act on it is
through the evaluation at some point.
The elements of $T^{*}\hat \lieg$ of
our interest are the pairs:
\eqn\prs{P(x),k\p_{x} + Q(x)} where $k$ is a
fixed number, $P(x)$
is a $\lieg$-valued function on a circle ${\bf S}^{1}$
and $Q(x)$ is a gauge field
on a circle. The phase space is acted on by the
gauge group:
\eqn\ggr{P(x) \mapsto g(x)^{-1}P(x) g(x), \quad Q(x) \mapsto
g(x)^{-1}Q(x) g(x) +
k g^{-1}(x) \p_{x} g(x)}
The moment equation has the
form:
\eqn\mmmtcm{
k\p_{x} P + [ Q, P] = J \delta(x)}
where $J$ is the one
from \mmprcm.
The number $k$ can be rescaled by the choice of the radius of
a circle $\bf S^{1}$. Instead
we choose the circle of unit radius and keep
$k$.
To solve the equation \mmmtcm\ we fix a gauge \ggr. We can either decide
that  $Q$
is a constant diagonal matrix
$$
Q = {\rm diag}(q_{1}, \ldots,
q_{N})
$$
and then the solution for $P(x)$ will produce the Lax operator
\lxtcm\ of the Sutherland
model with $R = {{2\pi}\over{k}}$ \gncal,\gnru.

It
is quite amusing that the same reduction yields the rational
$RS$ model as
well.
In order to see that choose the gauge
\eqn\fngge{
P(x) = {\rm diag}\left( p_{1}\left(x\right), \ldots, p_{N}\left(x\right)\right)
}
Then the moment equation \mmmtcm\ implies
that $Q(x)$ is diagonal everywhere
except $x=0$ where it has an
off-diagonal part proportional to the delta-function.
At the same time $P(x)$ is
forced to be constant $p_{i}(x) = q_{i}$.
\eqn\dltans{
Q(x)_{ij}  =  \theta_{i} (x)\delta_{ij} + {\delta} (x)
{{i\nu}\over{q_{i} - q_{j}} }
}
The natural candidate for a Hamiltonian in
this setting would be a gauge invariant function
of $Q(x)$. Since $Q(x)$ is
actually a gauge field the gauge invariant function
is a trace in some
representation $R$ of a Wilson loop:
\eqn\trcwl{
H_{R} = {\Tr}_{R} P\exp
\oint {1\over{k}} Q(x)dx
}
which is easy to evaluate provided we assume the
following structure
of the diagonal piece of $Q(x)$ (which supported by the
alternative derivation
of the solution to the moment equation
below):
$$
\theta_{i}(x) = \varphi_{i}(x) + \delta (x) \sum_{k \neq i}
{{i\nu}\over{q_{k}-q_{i}}}
$$
which makes the Wilson loop
\eqn\advc{
B =
P\exp \oint {1\over{k}} Q(x)dx = {\rm diag}\left(e^{{1\over{k}} \oint
\varphi_{i}(x)dx}\right)
\exp \left( {{i\nu}\over{k}} {\rm r} \right)
}
with
$\rm r$ being the matrix:
\eqn\rrm{{\rm r}_{ij} =
{1\over{q_{ij}}}, \quad i
\neq j, \quad {\rm r}_{ii} = - \sum_{j \neq i} {\rm r}_{ij}}
It is shown in the Appendix (in the trigonometric case from which this one
follows as well) that the matrix $B$ is gauge equivalent
to \lxrrstr\ with the
identification $\beta = {1\over{k}}$, and (cf. \lxrrstr):
$$
p_{i} +
{1\over{2i \beta}}
{\rm log} (-{\phi}^{+}_{i}/{\phi}^{-}_{i}) = \oint
\varphi_{i}(x)dx.$$

One can also get the same matrix (without the assumptions like
\dltans) $B$ by performing
a reduction of $T^{*}G$ under the adjoint action
of $G$ at the same level $J$
of the moment map:
$$
\mu = B^{-1} P B - P = J
$$

\noindent
\sssec{Trigonometric \quad RS = Relativistic
\quad trigonometric
\quad CM \quad model.} There are three different
approaches, all leading to
the same finite-dimensional
Hamitlonian system. There are two Hamiltonian
reductions and one Poisson reduction.
The advantage of Hamiltonian one  is
the simplicity and geometric clarity. The advantage
of Poisson one is
finite-dimensionality at each step and considerable simplicity of the
proof
of the canonical commutation relations. We try to outline all three
approaches with the emphasis
on the Poisson reduction, as the
relevant
Hamiltonian reduction was described in some details in \gnru. We
keep in mind a sequence of
contractions:
\eqn\sqvnc{
{\CA}_{{\bf T}^{2}}
\rightarrow T^{*}{\hat G} \rightarrow G \times G}
where the first entry is
the space of $G$-valued gauge fields on a two-torus ${\bf T}^2$,
the second entry is the
cotangent bundle to the central extension of the loop group
$LG$ and the last
one is the space of lattice (for the simplest graph, representing
a
two-torus) connections, described below.

\noindent
\sssec{Hamiltonian \quad approach.}
Consider the space ${\CA}_{{\bf T}^{2}}$ of
$SU(N)$ gauge fields $A$ on a two-torus
${\bf T}^{2} = {\bf S}^{1} \times {\bf S}^{1}$.
Let the circumferences of the circles
be $R$
and $\beta$.
The space  ${\CA}_{{\bf T}^{2}}$ is acted on by a gauge group $\CG$ ,
which preserves a symplectic form
\eqn\smplf{\Omega = {k\over{4{\pi}^{2}}}
\int {\Tr} \delta A \wedge \delta A,}
with $k$ being an arbitrary real
number for now.
The gauge group acts via evaluation at
some point $p \in {\bf T}^{2}$
on any coadjoint orbit $\CO$ of $G$, in particular, on $\CO =
{\IC\IP}^{N-1}$.
Let ${\IC\IP}^{N-1}$ have a $-N{\nu} \times$ Fubini-Study
symplectic form. Let $(e_{1} :  \dots : e_{N})$
be the homogeneous
coordinates on $\CO$. Then the moment map for the action of $\CG$
on $\CA_{{\bf T}^{2}} \times \CO$ is
\eqn\mmp{k F_{A} + J \delta^{2}(p),  \quad J_{ij}
= i{\nu} (  \delta_{ij} - e_{i} e_{j}^{*}) }
$F_{A}$ being the curvature
two-form. Here we think of $e_{i}$ as being the coordinates on $\IC^{N}$
constrained so that $\sum_{i} \vert e_{i} \vert^2  =N$ and
considered up to the multiplication
by a common phase factor.

Let us
provide a certain amount of commuting Hamiltonians. Obviously,
the
eigen-values of the monodromy of $A$ along any fixed loop on ${\bf T}^{2}$
commute
with themselves. We consider the reduction at the zero level of the moment
map.  We have at least $N-1$ functionally independent commuting
functions on
the reduced phase space $\CM_{\nu}$.

Let us estimate the dimension
of
$\CM_{\nu}$.
If $\nu= 0$ then the moment equation forces the connection to be
flat and therefore
its gauge orbits are parameterized by the conjugacy
classes of the monodromies
around two non-contractible cycles on ${\bf T}^{2}$: $A$
and $B$. Since the fundamental group $\pi_{1} ({\bf T}^{2})$ of ${\bf T}^{2}$
is abelian $A$ and $B$ are to
commute. Hence they are simultaneously diagonalizable, which makes
${\CM}_{0}$ a
$2(N-1)$ dimensional manifold. Notice that the generic point on the
quotient
space has a non-trivial stabilizer, isomorphic to the maximal torus
$T$ of $SU(N)$. Now, in the presence of $\CO$
the moment equation implies
that the connection $A$ is flat outside of $p$ and has a
non-trivial
monodromy around $p$. Thus:
\eqn\mmmeqt{ABA^{-1}B^{-1} = {\exp}
(R\beta J)}
(the factor $R\beta$ comes from the normalization of the
delta-function in \mmp).
If we diagonalize
 $A$, then $B$ is uniquely
reconstructed up to the right
multiplication by the elements of $T$. The
potential degrees of freedom in $J$ are
"eaten" up by the former stabilizer
$T$ of a flat connection: if we conjugate both
$A$ and
$B$ by an element $t
\in T$ then $J$ gets conjugated. Now, it is important that
$\CO$ has
dimension $2(N-1)$. The reduction of $\CO$ with respect to $T$ consists of a
point and does not contribute to the dimension of $\CM_{\nu}$. Thereby we
expect to get an integrable system. Without doing any computations we already
know
that we get a pair of dual systems. Indeed, we may choose as the set of
coordinates the eigen-values
of $A$ or the eigen-values of $B$. The logarithms of the
eigen-values of $B$ are the action
variables for the system generated by ${\Tr} B^{k}$, and vice versa.

The two-dimensional
picture has the advantage that the geometry of the problem
suggests the
$SL_{2}({\IZ})$-like duality. Consider the operations $S$ and $T$  realized
as:
\eqn\slgentrs{
S: (A, B) \mapsto (ABA^{-1}, A^{-1}); \quad T: (A,B)
\mapsto (A,  BA)}
which correspond to the freedom of  choice  of generators  in
the fundamental group of a two-torus. Notice that
both $S$ and $T$ preserve the commutator $ABA^{-1} B^{-1}$ and
commute with the action of the gauge group.
The group $\Gamma$ generated by $S$ and $T$ (it is a subrogup
of the group ${\bf OutFree} (2)$
of the outer authomorphismes of the
free group with two generators)
seems to be larger
then ${\rm SL}_{2}({\IZ})$. However in the limit $\beta, R \to 0 $ it
contracts to ${\rm SL}_{2}({\IZ})$ in a sense that
we get the
transformations
\gnsltwo\ by expanding
$$
A = 1 + \beta P + \ldots, \quad B = 1 + R Q +
\ldots
$$
for $R, \beta \to 0$.

The disadvantage of the two-dimensional
picture us the necessity to keep too many redundant
degrees of freedom. The
first of the contractions \sqvnc\ actually allows to replace the space of
two
dimensional gauge fields by the cotangent space to the (central extension of)
loop group:
$$
T^{*}{\hat G}   = \{ ( g(x), k\p_{x} + P(x) ) \}
$$
which is a
``deformation'' of the phase space of the previous example
($Q(x)$ got
promoted to a group-valued field). The relation to the two dimensional
construction
is the following. Choose a non-contractible circle $\bf S^{1}$
on the two-torus
which does not pass through the marked point $p$. Let $x,y$
be the coordinates on the torus
and $y=0$ is the equation of the $\bf S^{1}$.
The periodicity  of $x$ is $\beta$ and that of $y$ is $R$.
Then $$
P(x) = A_{x}(x,0),
g(x) =P\exp\int_{0}^{R} A_{y}(x,y) dy.
$$
The gauge transformations
on $\bf S^{1}$ transform  on $(g(x), P(x))$ is a way, similar to \ggr. The moment map equation
\mmp\
goes over to the moment map equation \gnru:
\eqn\mmppr{k g^{-1} \p_{x}
g + g^{-1}Pg - P = J \delta(x),} with $k = {1\over{R\beta}}$. The solution of
this equation in the gauge $P = {\rm diag}(q_{1}, \ldots, q_{N})$
leads to
the Lax operator $A = g(0)$ of the form \lxrtrstr\ with $R,\beta$ exchanged
\gnru. On the other hand,
if we follow \fngge\ and diagonalize
$g(x)$:
\eqn\fngget{g(x) =  {\rm diag} \left( z_{1} = e^{iR q_{1}}, \ldots, z_{N} = e^{iR q_{N}} \right)}
then a similar calculation
leads to the Lax operator
$$
B = P\exp\oint{1\over{k}} P(x)dx =  {\rm diag} (
e^{i \theta_{i} } ) \exp iR\beta\nu {\rm r}
$$
with
$$
{\rm r}_{ij} =
{1\over{1- e^{iRq_{ji}}}}, i\neq j; \quad {\rm r}_{ii} = - \sum_{j\neq i}
{\rm r}_{ij}
$$
thereby establishing the duality $A \leftrightarrow B$
explicitly.

\noindent
\sssec{Poisson \quad description.} Here we introduce a set of
commuting functions on the space of graph connection on a graph,
corresponding to a moduli space of flat connections on a torus with one hole
and describe the flow generated by this set. Being reduced to a particular
symplectic leaf of the  moduli space of flat connections on the torus , this
set of functions turns out to be a full set of commuting Hamiltonians. We
introduce another full set of commuting variables and write down the
Hamiltonians taking the latter set as a set of coordinates thus recovering
the Ruijsenaars integrable system.
Consider a graph, consisting of two
edges and one vertex with the fat graph structure
corresponding to a
punctured torus \fr.
%as shown on fig.1.
%\begin{equation}
%{\epsfxsize16\baselineskip\epsfbox{ruijenaars1.eps}}
%\end{equation}
%This  graph corresponds to the torus with one hole (fig. 2).
The space of graph
connections ${\cal A}^L$ for such graph is
just a product of two copies of
the group $G$: $ {\cal A}^L = G \times G = \{ (A,B) \vert A, B \in G \}$,
where $A$ and $B$ are
assigned to the edges of the graph.
%\begin{equation}
%{\epsfxsize16\baselineskip\epsfbox{ruijenaars2.eps}}
%\end{equation}
For a choice of ciliation on ${\cal A}^L$
%as shown on fig. 1.
the Poisson bracket on ${\cal A}^L$
is given by the relations, following from
the general rules \fr.
\eqn\psbrctks{\eqalign{
  \{ A {\ocom} A \} \quad =& \quad r^a A\otimes A +
A\otimes A r^a - 2 (A \otimes 1) r^a (1 \otimes A)\cr
  \{ B {\ocom} B\} \quad = & \quad r^a B\otimes B + B\otimes B r^a - 2
(B \otimes 1) r^a (1 \otimes B)\cr
 \{A {\ocom} B\} \quad = &
\quad r  (A\otimes B) + A \otimes B r + ( 1 \otimes B) r_{21} (A \otimes 1) -
(A \otimes 1) r (1 \otimes B),\cr}}
where $r^a = {\half}(r - r_{21})$.

Now let us restrict ourselves to the case $G={\rm SL}_{N}$ and
the standard $r$-matrix:
\eqn\rma{
 r = \sum_{\alpha > 0} E_{\alpha} \otimes E_{-\alpha} + \half
\sum_i H_i \otimes H_i, \quad r^a = {\half} \sum_{\alpha > 0} E_{\alpha} \wedge
E_{-\alpha} }
In this case one can easily derive the following commutaion relations
\eqn\comm{
\{ {\Tr} A^n, A\} = 0 \quad  \{ {\Tr} B^n, B\} = 0}
\eqn\bra{  \{ {\Tr} A^n, B \} = n(A^n)_0 \quad
  \{ {\Tr} B^n, A\}  = nA(B^n)_0}
where $(X)_0$ denotes the traceless part of the matrix $X$.
  Therefore, the functions ${\Tr} B^n$ for $n=1 \ldots N-1$ considered as Hamiltonians generate commuting flows on ${\cal A}^L$.
\eqn\eone{\eqalign{
   B\left( t_1, \ldots, t_{N-1} \right) & = B \left( 0,\ldots ,0\right) \cr
   A\left( t_1,\ldots,t_{N-1} \right) & = A
\left( 0,\ldots, 0\right)
e^{\bigl( t_1 B + \cdots + t_{N-1}B^{N-1}\bigr)_{0}}\cr}}
As it was shown in \fr\ the lattice gauge group ${\cal G}^L$ acts on ${\cal G}^L$ in a Poisson way, and the quotient Poisson manifold coincides with the moduli space ${\cal M}$ of smooth flat connection on the Riemann surface, corresponding to the fat graph $L$. In our case the group ${\cal G}^L$ is $G$ itself (for the graph has just one vertex) which acts on $A$ and $B$ by simultaneous conjugation.
\eqn\mpst{
 g: (A,B) \mapsto (gAg^{-1}, gBg^{-1}).}

The functions ${\Tr} A^{k}$ and ${\Tr} B^{k}$ are invariant under this action, and therefore their pull-downs on the moduli space ${\CM}$ generate commuting flows there, which trajectories are just projections of \eone.

However the moduli space ${\CM}$ in our case is a Poisson manifold with degenerate Poisson bracket. The Casimir functions of this Poisson structure are the functions of conjugacy classes of monodromies around holes and constant value levels of such functions are just the symplectic leaves of ${\CM}$.
In our case such Casimir functions are  ${\Tr} \bigl( ABA^{-1}B^{-1} \bigr)^{k}$, pulled down to ${\cal M}$.

Different symplectic
leaves have different dimensions
and the lowest dimension of them is $2(N-1)$. These leaves
correspond to the monodromy around the hole conjugated to a matrix
$$e^{-i R\beta\nu} {\Id} + P,$$
${\rm rk} P \leq 1$, $\nu$ is a numerical constant from the previous
section
parameterizing the set of symplectic leaves of lowest dimension.
Let $t = e^{-iR\beta\nu}$. On the leaf ${\CM}_{\nu}$
the family of functions
${\Tr} A^k, k=1, \ldots, N-1$ forms a full set of Poisson-commuting variables.

Introduce local coordinates on these symplectic leaves in the following way.
Let $z_1 = e^{i R q_{1}}, \ldots, z_N = e^{iR q_{N}}$
be the eigenvalues of the operator $A$ and $\mu_1, \ldots, \mu_N$
are the corresponding diagonal matrix elements of $B$ (in the basis, diagonalizing $A$). One can check that in this basis
\eqn\lxb{B^i_j = \sqrt{\mu_i \mu_j}{{(1-t)}\over{ z_i / z_j - t}}.}
The functions $z_i$ and $\mu_j$ are  well-defined
locally on the symplectic leaf ${\CM}_{\nu}$.
Their Poisson brackets are equal to:
\eqn\athree{\eqalign{
  \{z_i,z_j\} &= 0\cr
  \{\mu_i,\mu_j\} &= \mu_i \mu_j {{(z_i + z_j)}\over{(z_i / z_j - t)(z_j / z_i - t)(z_i - z_j)}} \quad i\neq j \cr
  \{z_i, \mu_j\} &= z_i \mu_j \delta_{i,j}.\cr}}

To define the variables, canonically conjugated to $z_i$ we can just multiply $\mu_i$ by  factors independent on $\mu_i$. For example one can take:
\eqn\hrn{
   s_i = \mu_i t^{{N-1}\over{2}} \prod_{k,k \neq i}
\sqrt{{(z_k-z_i)(z_i-z_k)}\over
{(z_k-tz_i)(z_i-tz_k)}}}
One can check, that these new variables $s_i$ have the Poisson brackets
\eqn\npsb{
\{s_i,s_j\} =0\quad
 \{z_i,s_j\} = z_i s_j \delta_{i,j}.}
Substituting this back to the formula  \lxb\  we get:
\eqn\lxoag{B_{ij} = {{1-t}\over{z_{i}/z_{j} - t}} \left( \Phi_{i}^{+}\Phi_{i}^{-}\Phi_{j}^{+}\Phi_{j}^{-} \right)^{1/4}}
which is gauge equivalent to \lxrtrstr.

\noindent
\sssec{Moral \quad revisited.} We have seen in all the previous examples that the
origin of the dual system is connected with the existence of transversal $G$-invariant
foliations on the original space, which become Lagrangian foliations when pulled
down to the quotient. The simplicity of the operating with dual systems in the advocated framework in the classical case allows one to hope that the
 duality can systematically elevated to
the quantum case as well. See \ch\ek.
\subsec{Appendix. Computation of the Poisson brackets}
The bivector defining the Poisson structure on ${\cal A}^L$ can be rewritten in the form
\eqn\bvctr{
 {\pi} = {\half} \sum_{i,j,u,v} E^{i (u)}_j \otimes E^{j (v)}_i (\epsilon(u,v) + \epsilon(i,j)),}
where  $\epsilon(i,j)$ is $-1, 0$ or $1$ depending on whether
 $i$ is less, equal or greater than $j$ respectively and
$E^{i (u)}_j$ are the standard ${\rm GL}_{N}$
generators acting on the $u$-th end of the edge.
(In our case  $E^{i (1)}_j$ acts on $A$ from the left,  $E^{i (2)}_j$ acts on $B$ from the left, $E^{i (3)}_j$ acts on $A$ from the right  and $E^{i (4)}_j$ acts on $B$ from the right.)

It is not convenient to
compute the Poisson brackets
between $ z_i$ and $\mu_j$ using this bivector directly,
for it does not preserve the diagonality of the matrix $A$.
However, to compute the Poisson brackets of
the gauge invariant functions we
are allowed to add to this bivector any
term vanishing on such functions i.e. any terms of the type
$\sum_u E^{i (u)}_j \otimes X$ or $X \otimes \sum_u E^{i (u)}_j $, where $X$ is an arbitrary vector field.
Using this and also the fact that
for the diagonal $A$:
\eqn\frdgnl{\eqalign{
&E^{i (1)}_j  = {{z_i}\over{z_j- z_i}}(E^{i (2)}_j +E^{i (4)}_j )\cr
&E^{i (3)}_j  = {{z_j}\over{z_i- z_j}}(E^{i (2)}_j +E^{i (4)}_j )\cr}}
the bivector $\pi$ can be transformed to the form:
\eqn\newbvct{
{\pi}^{\prime}= \sum_{i>j} E^{i (2)}_j \wedge E^{j(4)}_i
{{z_i + z_j}\over{2(z_i-z_j)} } +
\half \sum_{i} E^{i (2)}_i \wedge E^{i(1)}_i + E^{i (3)}_i \wedge E^{i(4)}_i}
Applying this bivector for the chosen $A$ and $B$ we get the desired Poisson brackets. $\flat$

The equations \frdgnl\  are the infinitesimal forms of
the statements that (i)
up to the gauge transformation the conjugation of $A$ by
$g$ is equivalent to the conjugation of $B$ by $g^{-1}$ thanks to \mpst\ \semi
(ii)
$g_{L} A = A^{-1} ( A g_{R}^{-1} ) A$, with $g_{L} = g_{R}^{-1}$.
\subsec{Appendix. Solution of the moment equation}

Here we solve the equation
 \mmmeqt:
\eqn\trmom{\eqalign{& A^{-1}BAB^{-1} = \exp R \beta J \cr
& J = - i \nu ( \Id - e \otimes e^{\dagger}), \quad \langle e^{\dagger}, e \rangle = N \cr}}
with  $A$, $B$ - $N \times N$ unitary matrices defined up to the gauge transformations
\mpst. We use the notation: $\alpha = R \beta\nu$.
We partially fix a gauge:
\eqn\dgge{A = {\rm diag} \left( e^{iRq_{1}} , \ldots, e^{iRq_{N}} \right)}
which leaves gauge transformations of the form
\eqn\dgge{h = \exp \left( i \, {\rm diag}(l_{1}, \ldots, l_{N}) \right).}
which preserve $A$, conjugate $B$ and map $e$ to $h^{-1}e$.
The exponent $\exp R\beta J$ is easy to compute:
$$
\exp  J = e^{-i\alpha} \left(
{\Id} + {{e^{iN\alpha }-1}\over{N}} e \otimes e^{\dagger} \right)
$$
Let $f = B^{-1}e$, $z_{i} = e^{iRq_{i}}$, $\Phi_{i}^{+} := \vert e_{i} \vert^{2} $, $\Phi^{-}_{i} := \vert f_{i} \vert^{2} $. Then:
\eqn\bij{\eqalign{& B_{ij} = e^{-i\alpha}{{e^{iN\alpha}-1}\over{N}}
 {{e_{i}f_{j}^{*}}\over{e^{iRq_{ji}}- e^{-i\alpha}}} \cr
& f = B^{-1}e \Rightarrow {{Ne^{i\alpha}}\over{e^{iN\alpha}-1}} =
\sum_{i=1}^{N} {z_{i}{\Phi^{+}_{i}}\over{z_{j} - e^{-i\alpha} z_{i}}} \cr}}
The last equation implies (see below):
\eqn\fiplus{\Phi_{i}^{+} = {{N}\over{e^{-iN\alpha}-1}}
{{P(e^{-i\alpha}z_{i})}\over{z_{i} P^{\prime}(z_{i})}}, \quad
P(z) = \prod_{i=1}^{N} (z - z_{i}) }
Now the unitarity of $B$ implies, that
\eqn\unb{
\delta_{ik} = f_{i}f_{k}^{*}  {{e^{iN\alpha}-1}\over{N}} \sum_{j=1}
{{P(e^{-i\alpha}z_{j})}\over{z_{j}P^{\prime}(z_{j})
(z_{j}/z_{i} - e^{i\alpha})
(z_{k}/z_{j} - e^{-i\alpha})}} }
Hence
\eqn\fiminus{\Phi_{i}^{-} = {{N}\over{e^{iN\alpha}-1}}
{{P(e^{i\alpha}z_{i})}\over{z_{i}P^{\prime}(z_{i})}}}

To prove  \fiplus\ consider the  contour integral
$$
{1\over{2\pi i}} \oint_{\IGa} {{P(e^{-i\alpha} z)dz}\over{P(z)
( z - e^{i\alpha} z_{j})}}   = {\rm Res}_{\infty} = e^{-iN\alpha}
$$
To prove \fiminus\ consider the integral:
$$
{1\over{2\pi i}} \oint_{\IGa} {{P(e^{-i\alpha} z)dz}\over{P(z)
( z - e^{i\alpha} z_{i}) ( z_{k}  - e^{-i\alpha} z) }}
= \delta_{ik} {\rm Res}_{e^{i\alpha} z_{k}} =
\delta_{ik}  {{P^{\prime}(z_{k})}\over{P(e^{i\alpha} z_{k})}}
$$
In both cases the contour $\IGa$ surrounds the roots of $P(z)$.

Notice that both $\Phi_{i}^{\pm}$ are real:
\eqn\fipm{\Phi_{i}^{\pm} =
{{N {\rm sin}({\alpha}/2)}\over{{\rm sin} (N{\alpha}/2)}} \prod_{j \neq i}
{{{\rm sin} \left( {{Rq_{ij} \pm \alpha}\over{2}} \right)}\over{{\rm sin} \left( {{Rq_{ij}}\over{2}} \right)}}
}
Substituting this back to \bij\ we get:
\eqn\bbij{B_{ij} = e^{i \left( (1-N){\alpha}/2 + Rq_{ij}/2 + \varepsilon_{i} - \varphi_{j}\right)} {{{\rm sin}\left({{N{\alpha}}\over{2}}\right)}\over{N {\rm sin} \left( {{Rq_{ij} + {\alpha}}\over{2}} \right)}}
{\sqrt{\Phi_{i}^{+} \Phi_{j}^{-}}} }
where $e_{i} =: \vert e_{i} \vert e^{i\varepsilon_{i}}$, $f_{j} =: \vert f_{j} \vert e^{i\varphi_{j}}$. The
gauge transformations \dgge\ allow us to set $\varphi_{i} + Rq_{i}/2= 0$. Then define
\eqn\cnj{p_{i} = - {1\over{\beta}}
\left(  (1-N){\alpha}/2 + Rq_{i}/2 + \varepsilon_{i} \right)}

Finally, the matrix $B$ can also be written as:
\eqn\exfrm{B = \left( \Phi^{-} \right)^{-\half}
\left( e^{-i\beta\tilde p} e^{-i\alpha \rm r} \right) \left( \Phi^{-} \right)^{\half}}
where
$\Phi^{-} = {\rm diag} ( \Phi^{-}_{i} )$,
$\tilde p = {\rm diag} ({\tilde p_{i}} )$,
\eqn\shftp{\tilde p_{i} = p_{i} - {{\alpha + \pi}\over{2\beta}} + {i\over{2\beta}} {\rm log} \left(
{{\Phi^{-}_{i}}\over{\Phi^{+}_{i}}} \right) }
and
\eqn\smlrtr{\eqalign{&{\rm r}_{ij} = {{z_{i}}\over{z_{i}- z_{j}}},
\quad i \neq j \cr
& {\rm r}_{ii} = {\half} {{z_{i} P^{\prime\prime}(z_{i})}\over{P^{\prime}(z_{i})}} \cr}}

To prove the last statement consider the matrix
$$
R_{ij}(\alpha) = {{\sin\left( {{N\alpha}\over{2}} \right)}\over{N\sin\left( {{Rq_{ij} + \alpha}\over{2}} \right)}}
\phi_{i}^{+} = e^{{i(N-1)\alpha}\over 2} \sqrt{{z_{j}}\over{z_{i}}} {{P(e^{-i\alpha} z_{i})}\over{(e^{-i\alpha} z_{i} - z_{j})P^{\prime}(z_{i})}}
$$
We have:
$$
B = \left( \Phi^{-} \right)^{-\half} \left( e^{-i\beta\tilde p} R(\alpha)
\right) \left( \Phi^{-} \right)^{\half}
$$
A simple contour integral calculation shows that
$$
R (\alpha_{1}) R(\alpha_{2}) = R( \alpha_{1} + \alpha_{2})
$$
The rest follows by expanding near $\alpha=0$.
If one performs an expansion near $R = 0$
one gets the statement that  the rational Lax operator
\lxrrstr\ is conjugated to the operator of the form announced in \advc.

It is amusing that the expression $ {1\over{2i\beta}}
{\rm log} ({\Phi}^{+}_{i}/{\Phi}^{-}_{i}) $ appears quite often in
Bethe Ansatz Equations
for $XXX$ magnets and their field theoretic limits
\fadba.
\vfill\eject
%%%%%%% Konec glavy 4 %%%%%%%%

%% file: dual5.tex
%%%%%%% Glava 5 %%%%%%%
%%%%%%%%%%%%%%%%%%
%%%%%%% Fizika %%%%%%%

\newsec{Gauge theories and duality in integrable systems}
\subsec{Old approach: many-body systems as low-dimensional gauge theories}
It is a fruitful approach to think of the many-body system as of the gauge theory of
a certain kind. Namely, the particles of the model can be identified (sometimes)
with the eigenvalues of the Wilson loops in the theory and the gauge dynamics
becomes a dynamics of the particles. Of course, in the real
four dimensional world the gauge field has infinitely many degrees
of freedom and we don't expect to see any tractable quantum mechanical
system unless we have a principle which allows us to restrict
the dynamical problem to a finite number of degrees of freedom. The simplest
case is the case of low-dimensional gauge theory, where
the gauge field simply doesn't have propagating degrees of freedom.
Consider, for example, two dimensional Yang-Mills theory with a
gauge group $G = U(N)$. When formulated on a circle of radius $\bf R$
in the Hamiltonian formalism
the theory has as a phase space the space of gauge fields $A(x)$
on the circle and
their duals - chromoelectric fields $E(x)$. The gauge group acts on $(E,A)$ as
follows:
\eqn\ggrym{(E, A) \to (g^{-1}E g,  g^{-1} \p_{x} g + g^{-1}Ag)}
leading to the Gauss law $\p_{x} E + [A, E]$, which is nothing
but the moment map from the section $4.2$. We can go to the gauge where
$A$ is a constant (w.r.t. $x$) diagonal matrix
\eqn\gge{A = {\rm diag}( q_{1}, \dots, q_{N})}
Here are our particles. The time evolution makes $q_{i}$ to move and depending
on the circumstances such as the presence of the sources like $J$
(which correspond to the time-like Wilson lines) one gets the Hamiltionian system
of the kind we described and studied. The large gauge transformations
shift
$q_{i
}$'s by integer multiples of ${{1}\over{\bf R}}$ making
them live on a circle of radius ${1\over{2\pi \bf R}}$.
One can get more complicated examples by deforming
the model as follows. Replace $\bf S^{1}$ by $\bf T^{2}$, $E$ by the second
component of the gauge field along the torus, the symplectic
form being $\int_{\bf T^{2}} {\Tr} \delta A \wedge \delta A$. Then the Gauss law
becomes $F_{A} = \p_{x} A_{y} - \p_{y}A_{x} + [A_{x}, A_{y}]$.
Setting it to zero allows to diagonalize $A_{x}, A_{y}$ simultaneously:
\eqn\ggei{\pmatrix{A_{x}\cr A_{y}\cr} = {\rm diag} \left(
\pmatrix{p_{1}\cr q_{1}\cr}, \dots,
\pmatrix{p_{N}\cr q_{N}\cr}\right)}
Here, $x_{i}$ and $y_{i}$ do not Poisson-commute, although both live
on circle. One gets, therefore a system of relativistic partcles on a circle. The radius
of the circle is ${1\over{2\pi {\bf R}_{x}}}$, the speed of light is ${\bf R}_{y}$.
The gauge theory this model corresponds to is known as
Chern-Simons theory on a torus (perhaps with punctures).

One can go higher in dimensions with some care.
For example, by considering
supersymmetric $\CN=2$ theory in $d=4$ with compact space
$M^{3}$ one gets a quantum
mechanics on the moduli space of monopoles in ${\IR}^{3}$.

\subsec{New approach: many-body systems in supersymmetric gauge theories}

Recent progress in the understanding of
non-perturbative phenomena emerged after the work
of Seiberg and Witten on four dimensional $\CN=2$ SYM  \SeWi\
and works of Seiberg and his collaborators on $\CN=1$ $d=4$ theories.
The major tool in these studies is the low energy
effective Lagrangian which is constrained by two
priniciples - the holomorphy of chiral objects and
electric-magnetic duality. It is the electric-magnetic
duality which makes the integrable systems to appear in
the solutions to the gauge theories.

In particular, one can argue on the general grounds \WitDonagi\ that any $\CN=2$ supersymmetric
gauge theory in four dimensions corresponds to a certain
integrable system in the holomorphic sense. The point is that the Coulomb
branch of
the theory parameterizes the family of abelian varieties (whose period
matrix coincides with the matrix of  coupling constants of the
effective low-energy abelian
theory). Moreover the total space must carry a holomorphic symplectic form $\omega$,
whose integral along the cycle in the fiber gives rise to a derivative of the
central charge of a $BPS$ representation of $\CN=2$ susy algebra along
the base.
Moreover the abelian varieties must be Lagrangian with respect to $\omega$.

The integrable systems corresponding to a large number of field
theories are identified. In particular, the low-energy theory of the
pure $N=2$ $SU(N_{c})$ SYM
is governed by the $A_{N_{c}-1}$ periodic (or affine) Toda system.
The $\CN=2$ theory with a massive adjoint hypermultiplet
corresponds to the elliptic Calogero-Moser system,
where the mass (which is naturally
a complex parameter in the $\CN=2$ theory) is identified with the coupling constant.
The theory
is UV finite (in fact, it is softly broken $\CN=4$ theory) and
therefore has as another
modulus -- the ultra-violet coupling ${\tau}$
which enters the integrable model as the modulus of the curve.
Another theories which were
mentioned so far are the relativistic generalizations of those
two. These correspond to five dimensional gauge theories with the same
number of supercharges, compactified on a circle of a finite
radius $R$. The speed of light of the relativistic model is proportional to the inverse radius
${1\over{R}}$ of the circle. For the theories with fundamental
matter
the firm identification with the integrable
systems has been made in four \gmmra\ggmone\ as well
as in five and six dimensions \ggmone.

In some cases the dualities suggested by the integrable systems
are not obvious on the field theory side. We plan to return to
more detailed treatment of these cases (which involve six dimensional
theories) in the future.

\subsec{Dualities in field theories vs. dualities in many-body systems}

\noindent\sssec{Dualities \quad in \quad the \quad old \quad approach.}
Let us start with the two-dimensional Yang-Mills theory
with the gauge coupling $g^{2}$ formulated on a
Riemann surface of area $A$. It was shown
by E.~Witten in \Witdgt\ that the perturbative in $g^{2}A$ part of
the correlation functions in this (non-supersymmetric) theory coincides
with the correlation functions of certain observables in
twisted ${\CN}=2$ supersymmetric two-dimensional  Yang-Mills theory.

Among the twisted supercharges of the latter theory one finds a scalar
$Q$ which annihilates the complex scalar $\phi$ in the vector
multiplet. The observables constructed out of the gauge invariant
functions of $\phi$ and their descendants can be mapped
to certain observables in non-supersymmetric theory.

As we discussed above, when Yang-Mills theory is formulated on a cylinder
with the insertion of an appropriate time-like Wilson line, it
is equivalent to the Sutherland model describing a collection
of $N$ particles on a circle. The observables ${\Tr} \phi^k$
of the previous paragraph are precisely the integrals of motion of
this system.

One can look at other supercharges as well. In particular,
when the theory is formulated on a cylinder there is another
class of observables annihilated by a supercharge. One can arrange the
combination
of supercharges which  will annihilate the Wilson loop operator. By repeating
the procedure similar to the one in \Witdgt\ one arrives at the
quantum mechanical theory whose Hamiltonians are generated
by the spatial Wilson loops. This model is nothing
but the rational Ruijsenaars-Schneider many-body system.

The duality between these two systems is a consequence of the fact that
when lifted to the supersymmetric model both field theories
become equivalent to the same ${\CN}=2$ super-Yang-Mills
theory in two dimensions.

The self-duality of trigonometric Ruijsenaars system has even
more transparent physical meaning. Namely, the field theory
whose quantum mechanical avatar is the Ruijsenaars system is
three dimensional Chern-Simons theory on
${\bf T}^{2} \times {\bf R}^{1}$ with the insertion of an appropriate
temporal Wilson line and spatial Wilson loop. It is the freedom
to place the latter which leads to several equivalent theories.
The group of (self-)dualities of this model is very big
and is generated by the transformations $S$ and $T$ \slgentrs.

In short, the duality reveals here itself  as a consequence of
Lorentz invariance of the underlying field theory.

\noindent\sssec{Duality \quad in \quad the \quad new \quad approach.}
The new approach deals with supersymmetric gauge theories
in three, four, five and six dimensions. Perhaps the richest
case is the six dimensional theory compactified on a three dimensional
torus ${\bf T}^{3}$ down to three dimensions.

As was discussed extensively in \witseiii\
in case where two out of three radii of ${\bf T}^{3}$ are much smaller
then the third one ${\bf R}$ the
effective three dimensional
theory is a sigma model with the target space ${\CX}$ being the
hyper-kahler manifold (in particular, holomorphic symplectic)
which is a total space of algebraic integrable system. The complex structure
in which ${\CX}$ is the algebraic integrable system
  is independent of the radius ${\bf R}$ while the K\"ahler structure
depends on $\bf R$ in such a way that the K\"ahler class of the
abelian fiber is proportional to $1/{\bf R}$.

The duality of the integrable systems
shows up in the gauge theories in the  several ways.

First of all {\sl AA} duality the well-known phenomenon
in the four dimensional ${\CN}=2$ gauge theory  which was
observed and exploited in \SeWi\ and then later on in the
plenty of works. The low-energy effective theory has
different sets of relevant degrees of freedom over different
regions of the moduli space of vacua. The transformations
between different descriptions  go through the
electric-magnetic duality on the gauge field side which is
accompanied by supersymmetry by
a {\sl AA}-type duality on the scalar side. Although this
duality is connected with the electric-magnetic
symmetry which is not realized geometrically in four dimensions,
it does become geometric when the theory is lifted
to a tensor theory in six dimensions \witdynii.

The duality of the {\sl AC} type is also present and is rather
interesting.
As one varies the moduli of ${\bf T}^3$ the geometry of ${\CX}$ varies as well.
In particular, different four dimensional theories can flow to
the same three dimensional theory. This is  where the {\sl AC}
duality in the integrable systems shows up.

For example, a certain scaling
limit of the five dimensional $SU(N)$ theory with massive adjoint
hypermultiplet, compactified on a circle
seems to be equivalent/dual to  four dimensional
$SU(N)$ theory with massive adjoint hypermultiplet when instanton corrections
are turned off in both theories.

The theory in three dimensions which came from four dimensions
upon a compactification on a circle whose low-energy
effective action describes only abelian degrees of freedom
can be always dualized to the theory of scalars/spinors only,
due to the vector-scalar duality in three dimensions.
In this way different sets of vector and hypermultiplets
in four dimensions can lead to the same three dimensional
theory (one of the examples of such symmetries is
provided by the three dimensional mirror symmetry \Seiii\Kapiii).

One can also use the {\sl AC} duality to establish the
following fact. Consider two {\sl AC} dual integrable systems.
Consider three dimensional ${\CN}=4$ supersymmetric
gauge theory whose Higgs branch is ${\CX}$ - the phase space
of these systems. Then two ${\CN}=2$ supersymmetric
theories whose spaces of scalars are both ${\CX}$ but the
superpotentials are taken from the sets of Hamiltonians of the
first and the second systems respectively are
dual to each other in the sense that both flow
in the UV/IR (depending on the whether these Hamiltonians
correspond to the relevant or irrelevant operators)
to the same theory.

We are certain that there are more applications of
the notion of duality in integrable systems both in
the theory of integrability itself and in the physics, gauge theories
being the arena for the most immediate ones.